\documentclass[sigconf]{acmart}

\newcommand{\tool}{{\textsc{{Dango}}}}

\newcommand{\new}[1]{{\color{black}{#1}}}

\usepackage{xcolor}
\usepackage{soul}
\usepackage{colortbl}
\usepackage{framed}
\usepackage{color}
\usepackage{booktabs}
\usepackage{multirow} 
\usepackage{wrapfig}
\usepackage{caption}
\usepackage{subcaption}
\usepackage{fancybox}  
\usepackage{textcomp}
\usepackage{enumitem}
\usepackage{acmart-taps}

\definecolor{academicgray}{rgb}{0.7, 0.7, 0.7}

\aptLtoXcmd{}{
}

\aptLtoXcmd{}{

\newcommand{\clickablesection}[1]{\hyperref[#1]{Section~\ref{#1}}}}

\definecolor{lightgray}{rgb}{0.95,0.95,0.95}
\definecolor{framecolor}{rgb}{0.8,0.8,0.8}
\definecolor{bordercolor}{rgb}{0.8,0.8,0.8}  

\aptLtoXcmd{}{\renewenvironment{framed}{%
  \vspace{-0.2\baselineskip}
  \MakeFramed {\advance\hsize-\width \FrameRestore}}%
 {\endMakeFramed\vspace{-0.2\baselineskip}}
}

\newcommand{\mycircle}[1]{{\large 
\textcircled{\small #1}}}

\AtBeginDocument{%
  }

\copyrightyear{2025}
\acmYear{2025}
\setcopyright{cc}
\setcctype{by}
\acmConference[CHI '25]{CHI Conference on Human Factors in Computing Systems}{April 26-May 1, 2025}{Yokohama, Japan}
\acmBooktitle{CHI Conference on Human Factors in Computing Systems (CHI '25), April 26-May 1, 2025, Yokohama, Japan}
\acmDOI{10.1145/3706598.3714135}
\acmISBN{979-8-4007-1394-1/25/04}

\begin{document}

\title{\tool: A Mixed-Initiative Data Wrangling System using Large Language Model}

\author{Wei-Hao Chen}
\email{chen4129@purdue.edu}
\orcid{0009-0003-9108-8473}
\affiliation{%
  \institution{Purdue University}
  \city{West Lafayette}
  \state{Indiana}
  \country{USA}
}

\author{Weixi Tong}
\email{weixitong@hust.edu.cn}
\orcid{0009-0001-2318-3191}
\affiliation{%
  \institution{Huazhong University of Science and Technology}
  \city{Wuhan City}
  \state{Hubei}
  \country{China}
}

\author{Amanda Case}
\email{amanda-case@uiowa.edu}
\orcid{0000-0001-7027-7871}
\affiliation{%
  \institution{University of Iowa}
  \city{Iowa City}
  \state{Iowa}
  \country{USA}
}

\author{Tianyi Zhang}
\email{tianyi@purdue.edu}
\orcid{0000-0002-5468-9347}
\affiliation{%
  \institution{Purdue University}
  \city{West Lafayette}
  \state{Indiana}
  \country{USA}
  \postcode{43017-6221}
}

\begin{abstract}


Data wrangling is a time-consuming and challenging task in a data science pipeline. While many tools have been proposed to automate or facilitate data wrangling, they often misinterpret user intent, especially in complex tasks. We propose {\tool}, a mixed-initiative multi-agent system for data wrangling. Compared to existing tools, {\tool} enhances user communication of intent by: (1) allowing users to demonstrate on multiple tables and use natural language prompts in a conversation interface, (2) enabling users to clarify their intent by answering LLM-posed multiple-choice clarification questions, and (3) providing multiple forms of feedback such as step-by-step NL explanations and data provenance to help users evaluate the data wrangling scripts. We conducted a within-subjects user study (n=38) and demonstrated that {\tool}'s features can significantly improve intent clarification, accuracy, and efficiency in data wrangling. Furthermore, we demonstrated the generalizability of {\tool} by applying it to a broader set of data wrangling
tasks.
\end{abstract}



\begin{teaserfigure}
\centering
  \includegraphics[width=0.80\textwidth]{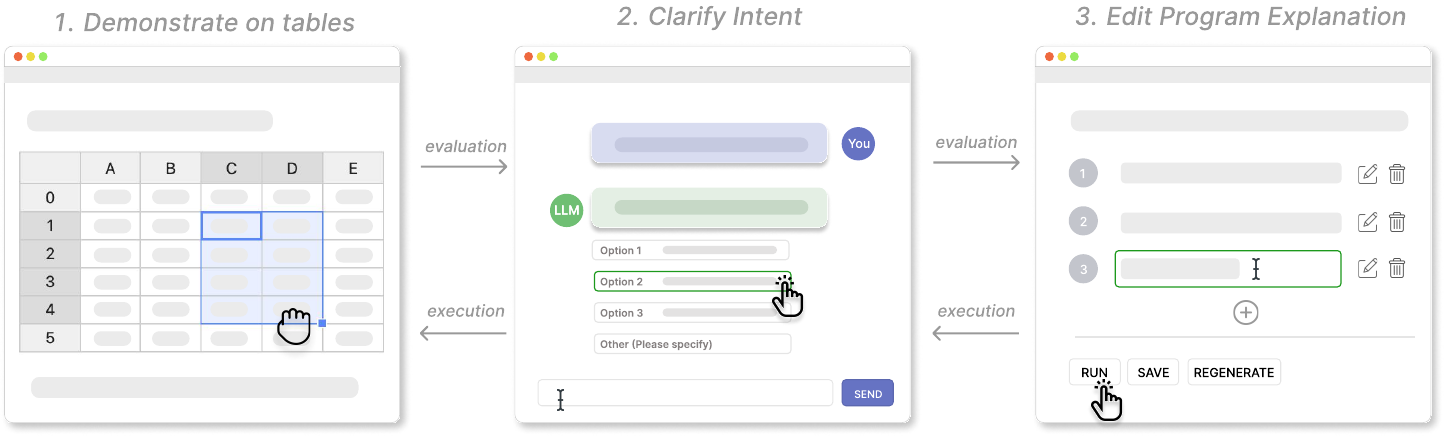}
  \caption{To create a data wrangling script, users can first demonstrate their desired actions on the tables in {\tool} (\textit{Step1}). They can edit table cells, add/delete/move columns and rows, and copy/cut content across multiple tables. For complex demonstrations, users can also describe their intent in natural language in a chatroom. \new{When ambiguity is detected in a demonstration or a NL description, {\tool}} will generate a multiple-choice question about each unclear part and prompt users for clarification (\textit{Step2}). Once the ambiguity is resolved, {\tool} synthesizes a data wrangling script to automate the desired actions. To make it easier for users to understand and validate the synthesized script, {\tool} explains the script in natural language step by step (\textit{Step3}). When users notice a wrong step in the script, they can easily fix it by directly editing the NL explanation of that step. They can also add missing steps or delete redundant steps. {\tool} will update the script based on user edits. 
}
  \label{fig:teaser}
\end{teaserfigure}

\maketitle


\section{INTRODUCTION}
Data wrangling is a time-consuming and challenging task in the early stages of a data science pipeline~\cite{biswas2022art, press2016cleaning}. It is reported that data scientists spend up to 80\% of their time changing table layouts, transforming data formats, and filling in missing or incorrect values~\cite{muller2019data}. 
To address this challenge, many interactive systems have been developed to help users clean data~\cite{kandel2011wrangler, jin2017foofah, gulwani2012spreadsheet, gulwani2011synthesizing}. However, they have two major limitations. First, they are limited to single-table tasks, while a recent study by Kasica et al.~\cite{kasica2020table} has shown that many data wrangling tasks involve multiple tables. Second, existing systems primarily rely on user demonstration to interpret user intent, in which the user demonstrates desired data transformations on a few data points and the system generates a generalizable script for automation. However, demonstration is not convenient to specify certain transformations, such as deleting a row only if more than a quarter of the values are missing. It is difficult to precisely communicate the criterion ``more than a quarter of the values are missing'' via demonstration alone.

The rise of Large Language Models (LLMs) provides new opportunities for augmenting or re-designing data wrangling tools. LLM systems have shown their ability to engage in natural language (NL) conversation with humans~\cite{achiam2023gpt, floridi2020gpt}, lowering the entry points for novice users to interact with more complex systems. More recently, several systems have been proposed to leverage LLMs for data wrangling~\cite{li2024sheetcopilot, srinivasa2022gridbook, chen2024sheetagent}. Despite their potential, these systems have several usability issues. First, they only allow NL interaction. While convenient, NL is inherently ambiguous and can be cumbersome when specifying certain operations compared with demonstrations, such as moving a specific row from one table to a target position in another
table. Second, these systems do not handle the hallucination problem~\cite{ji2023survey} and provide little support for users, especially end-users, in identifying and rectifying the errors in LLM-generated data transformations. Moreover, though LLMs are likely to misinterpret user intent, these systems do not provide effective ways for users to clarify their intent or provide feedback.

To address these issues, we propose {\tool}, a mixed-initiative system that enhances the communication between LLM agents and users for data wrangling. {\tool} supports both direct demonstration and NL interaction to enable rich intent expression. 
When user intent is ambiguous, {\tool} generates clarification questions (CQs). Users can refine their intent by answering these CQs. The Q\&A history is then incorporated into a feedback loop to continuously improve the results. To help users understand the script's behavior, {\tool} renders the synthesized scripts as step-by-step explanations in NL. Users can read these NL explanations easily and make direct edits to them. Given the complexity of multi-table tasks, {\tool} automatically visualizes data provenance, helping users understand the interrelationships between tables. 

We conducted a within-subjects user study with 38 participants to evaluate the usability and efficiency of {\tool}. Participants using {\tool} finished the assigned tasks in 3 min and 30 sec on average, reducing the task completion time by 32\% and 45\% compared with using the other two conditions. In the post-task survey, participants felt more confident about the generated scripts when using {\tool}. 
To understand the generalizability of {\tool}, we conducted a quantitative experiment on 24 additional data wrangling tasks. The result shows {\tool} can solve all of the tasks with an average task completion time of 62.67 and 75.90 seconds by the first two authors, respectively. This provides quantitative evidence about {\tool}'s effectiveness on a variety of data wrangling tasks.




\section{RELATED WORK}

\subsection{Data Wrangling}
Data wrangling is a critical but time-consuming process in data science. A recent study by Muller et al.~\cite{muller2019data} points out that even professional data scientists may take weeks, if not months, to wrangle data to achieve high data quality. This process has been reported to be one of the most time-consuming and laborious stages in a data science pipeline~\cite{dasu2003exploratory, kandel2011wrangler, press2016cleaning, muller2019data}.

To help end-users wrangle data, researchers have developed a variety of interactive data wrangling systems to reduce manual efforts. One of the earliest systems is Nix's Editing by Example system~\cite{nix1985editing}, which can automatically generate a replayable editing program by inferring users' input/output example text. Later, Witten and Mo~\cite{mo1992learning, witten1993tels} introduced the TELS system, which allows users to demonstrate text editing actions and uses these demonstration traces to synthesize text transformations. To improve the transparency of synthesized transformation scripts, Blackwell et al.~developed SWYN~\cite{blackwell2001swyn} that enables users to select text examples and preview the effects of the induced scripts. Similarly, Toped$^{++}$~\cite{scaffidi2009intelligently} allows users to create ``topes'' which are graphical representations of the data format used by the system to infer reformatting rules. Further developments such as  Potluck~\cite{huynh2007potluck} and Lapis~\cite{miller2001outlier} enable users to edit strings simultaneously at different locations. SMARTedit~\cite{lau2001learning} applied Programming-by-Demonstration (PBD) to automate text processing tasks by allowing users to demonstrate desired actions. Karma~\cite{tuchinda2008building} introduced a clean-by-example approach that allows users to specify the desired format of cleaned data. 

While the aforementioned systems focus on wrangling unstructured data such as text, a parallel line of research focuses on wrangling structured data such as tables or spreadsheets~\cite{ronen1989spreadsheet}. Wrangler~\cite{kandel2011wrangler, guo2011proactive}, for example, employed both PBD and direct manipulation to recommend applicable transformations based on user demonstrations on tables. Gulwani developed a new PBE algorithm~\cite{gulwani2011automating}, which is later used in Microsoft Excel known as FlashFill~\cite{gulwani2011automating}, allowing users to synthesize programs for string transformations. Harris and Gulwani~\cite{harris2011spreadsheet} later developed another PBE system that supports table structure transformation. To help data scientists wrangle data, Wrex~\cite{drosos2020wrex} applied PBE to synthesize wrangling scripts in Python within Jupyter Notebooks. Other researchers have also employed synthesis techniques to perform data transformations~\cite{yaghmazadeh2018automated, feng2017component, jin2017foofah, le2014flashextract, he2018transform, lau2001learning}. Despite their advancements, these systems only focus on synthesizing code for single-table tasks, while a recent study by Kasica et al.~\cite{kasica2020table} found that data wrangling tasks usually involve multiple tables. In contrast, {\tool} enables users to synthesize code for multi-table tasks.

The most related tools to our work are those that leverage LLM agents for wrangling tables and spreadsheets~\cite{li2024sheetcopilot, srinivasa2022gridbook, chen2024sheetagent, zhou2024llm, qi2024cleanagent, biester2024llmclean}. However, these systems only support one-shot synthesis and lack mechanisms for users to clarify or refine their intent, resulting in errors like incomplete solutions and wrong actions~\cite{li2024sheetcopilot}. This issue can be seen as the \textbf{gulf of execution}~\cite{hutchins1985direct}: \textit{the challenge users face in clearly expressing their intent to instruct AI.} To bridge this gap, {\tool} enables users to clarify their intent by answering AI-posed clarification questions before synthesis. Users can keep refining their intent through natural language instructions or demonstrations until the desired result is achieved.

\subsection{LLM-based Interactive System}
The recent advent of Large Language Models (LLMs), such as OpenAI’s GPT models~\cite{achiam2023gpt, floridi2020gpt}, Google's PaLM~\cite{anil2023palm}, and Meta's Llama~\cite{touvron2023llama}, has significantly influenced interactive systems across various domains~\cite{gao2024taxonomy}. For instance, Microsoft has integrated Copilot~\cite{microsoft2023copilot} into their suite of productivity tools. Amazon has developed a custom-built LLM for Alexa~\cite{amazon2023alexa}. In parallel, HCI researchers have also infused LLM into their system designs, revolutionizing fields in writing~\cite{chung2022talebrush, mirowski2023co, yuan2022wordcraft}, education~\cite{lieb2024student, kazemitabaar2024codeaid}, video editing~\cite{wang2024lave}, programming~\cite{mozannar2024reading, mcnutt2023design}, health~\cite{li2024conversational}, AR/VR~\cite{cheng2024scientific, liu2024classmeta}, visualization~\cite{wang2023data, vaithilingam2024dynavis}, data science~\cite{gu2024data, gu2024analysts, xie2024waitgpt}, among others. These LLM-based systems have introduced new interactive paradigms, shifting from traditional interactions—where users instruct computers \textbf{\textit{what to do}}—to a more AI-driven approach—where the user specifies \textbf{\textit{what results they want}}, a new interaction identified as \textit{intent-based outcome specification} by Nielsen~\cite{nielsen2023ai}.

However, recent studies show that these LLM-based systems could also misinterpret user intent due to the inherent ambiguity of NL instructions~\cite{gu2024analysts, liu2023wants, setlur2022you, danry2023don}, leading to misalignment between AI and human intent~\cite{shen2024towards}. To address these issues, many studies have used clarification questions (CQs) to resolve
ambiguous queries and improve user experience~\cite{aliannejadi2019asking, danry2023don, zou2023users, wang2021template, eberhart2022generating, tix2024better}. For instance, Danry et al.~\cite{danry2023don} found AI-framed Questioning increases human discernment accuracy of flawed statements. In the information retrieval community, CQs have been found useful to help user clarify their needs in search engines~\cite{zou2023users, aliannejadi2019asking, wang2021template}. Similarly, in the software engineering domain, Eberhart et al.~\cite{eberhart2022generating} introduced an approach for generating CQs to refine user queries in code search.

Despite its usefulness, clarifying ambiguous user intent using CQs has received limited attention in data wrangling. To the best of our knowledge, no data wrangling tools have utilized CQs to assist users in clarifying their intent. Traditional synthesis methods~\cite{kandel2011wrangler, gulwani2011automating, harris2011spreadsheet, drosos2020wrex}, such as PBD or PBE, typically lack effective mechanisms for refining user intent. In this work, {\tool} uses both PBD and a conversational interface powered by LLM to help users express their intent. When the user intent is unclear, {\tool} asks CQs to help refine their intent. In addition, {\tool} provides multiple feedback modalities, such as data provenance visualization and step-by-step program explanation, to help users evaluate the results, creating a rich feedback loop for effective data wrangling.



\section{DESIGN GOALS AND RATIONALE}

\subsection{Design Goals}
We reviewed prior work on data wrangling, especially those that have conducted user studies or discussed usability challenges in data wrangling tools~\cite{lau2008pbd, kasica2020table, kandel2011wrangler, drosos2020wrex, gulwani2012spreadsheet, gulwani2015inductive, li2024sheetcopilot, srinivasa2022gridbook, chen2024sheetagent, gu2024data}. We summarized four major design goals for {\tool} based on the common issues and challenges identified by previous work.

\textbf{G1. Help users express and clarify their intent}.
\label{sec:g1}
Previous tools~\cite{lau2008pbd, kandel2011wrangler, drosos2020wrex, gulwani2012spreadsheet} primarily use PBD to synthesize data wrangling scripts based on examples or demonstrations. However, many studies~\cite{lee2017towards, myers2001demonstrational, lau2009programming, zhang2020interactive} showed that users find it hard or cumbersome to provide complex examples. Recently, the rise of LLMs has brought new opportunities for facilitating data wrangling tasks with NL interfaces powered by LLM agents~\cite{li2024sheetcopilot, srinivasa2022gridbook, chen2024sheetagent, zhou2024llm, qi2024cleanagent, biester2024llmclean}. While this enriches the interaction modes, the inherent ambiguity of natural language also brings new challenges to interpreting user intent. According to a recent study by Gu et al.~\cite{gu2024analysts}, LLM agents could easily misinterpret user intent and generate incorrect operations. Moreover, instructing LLMs via natural language can be a brittle experience for non-AI-experts~\cite{zamfirescu2023johnny}. They often struggle with finding the right narratives to effectively communicate with LLMs. Therefore, it is critical to help users better express their intent and more importantly,  clarify their intent in case of misinterpretations.

\textbf{G2. Help users wrangle multiple tables}.
Previous data wrangling tools~\cite{lau2008pbd, kandel2011wrangler, drosos2020wrex, gulwani2012spreadsheet} only support single-table interactions. However, a recent study by Kasica et al.~\cite{kasica2020table} reveals that as data complexity and volume increase these days, data wrangling tasks often involve multi-table operations. Specifically, they identified 21 multi-table operations that are challenging to demonstrate and synthesize using existing synthesis methods. For instance, an auditor may need to split tables of annual transaction records over the past 20 years into subtables based on the product category and then only remove rows with missing values in each subtable. Currently, there is little tool support for such multi-table operations~\cite{kasica2020table}. Thus, the new tool should be able to interpret user demonstrations or NL descriptions that involve multi-table operations and synthesize scripts to perform these operations.

\textbf{G3. Help users interpret and validate the automation results.} Compared with other automation tasks, data wrangling faces a unique challenge due to the large volume of data involved in a task. It is hard and sometimes time-prohibitive to manually validate the wrangling results~\cite{olson2003data, gu2024analysts}. While PBD systems synthesize a program that prescribes the operations behind the wrangling results, it is difficult for end-users to understand the program, which is one of the main reasons that hinder the widespread adoption of PBD systems~\cite{lau2009programming, gulwani2015inductive}. Specifically, in the reflection paper on why PBD fails, Lau mentioned the system should ``\textit{encourage trust by presenting a user-friendly model}''~\cite{lau2009programming}. 
This echoes the \textit{understanding barrier} and the \textit{information barrier} in end-user programming~\cite{ko2004six}. 
More recently, Gu et al.~\cite{gu2024analysts} reveal the challenges of interpreting and validating data analysis results, such as not being able to translate the semantics of the analysis code to familiar operations in their mental models. Therefore, it is critical to design effective mechanisms to help users quickly interpret and validate the results or the code that computes the automation results. 

\textbf{G4. Help users make corrections.} For PBD systems, the defacto way of fixing an incorrect script is to provide more demonstrations to refute the incorrect script so that the PBD system can better extrapolate and avoid overfitting. However, this is inefficient, especially for minor errors, as the users must demonstrate from the beginning every time. Several studies have shown that end-users prefer simpler and more efficient error correction methods~\cite{hess2022informing, lau2008pbd}. Wrangler~\cite{kandel2011wrangler} and Wrex~\cite{drosos2020wrex} address this by enabling users to directly manipulate the synthesized script. However, Wrangler is limited to simple scripts, and Wrex still requires program comprehension and thus is not suitable for end-users. More recent LLM-based data wrangling systems~\cite{li2024sheetcopilot, chen2024sheetagent} allow users to make corrections by refining the initial prompt or providing NL feedback in the conversation. While this eliminates the need for directly modifying the synthesized script, users still need to overcome the communication barriers to LLMs~\cite{dang2022prompt, zamfirescu2023johnny}. In particular, a recent study~\cite{zamfirescu2023johnny} shows that non-AI-experts often struggle with finding the right prompt to effectively steer the output of an LLM. These challenges highlight the need to help users make corrections more efficiently, especially for LLM-based data wrangling tools.



\subsection{Design Rationale}
We made two specific design choices to help users to express and clarify their intent (\textbf{G1}). First, {\tool} allows users to express their intent using a combination of demonstration and natural language conversations. Both interaction modalities are convenient for end-users without a steep learning curve. Moreover, they are complementary to each other. Some actions, such as editing a specific cell and moving a row from one table to a specific position in another table, are easy to demonstrate by directly performing them on the table, while some actions are hard to demonstrate but easier to describe in natural language, such as specifying conditions for an action or performing a statistical test.  

Second, {\tool} leverages the intrinsic reasoning capability of LLMs to proactively detect potential ambiguity in user input and ask clarification questions. This proactive interaction design is motivated by previous findings that users often struggle with identifying which parts of their input lead to the erroneous output and how to effectively clarify their intent~\cite{gu2024analysts, zamfirescu2023johnny, liu2023wants}. Furthermore, when users prefer or find it necessary, they can still clarify their intent by directly providing additional demonstrations or NL feedback in {\tool}. This allows flexibility and forms a mixed-initiative interaction experience where the human and the agent take turns to contribute at the perceived appropriate time~\cite{allen1999mixed}.   


To achieve \textbf{G2}, we extended a popular DSL for single-table data wrangling~\cite{raman2001potter, kandel2011wrangler} to support multi-table operations, as detailed in Appendix~\ref{appendix:dsl}. We chose to synthesize scripts in a DSL rather than a general-purpose programming language such as Python, since the DSL has a much simpler grammar with highly abstracted operators, which makes it easier to synthesize and explain. Furthermore, we chose to leverage an LLM to synthesize a script instead of using traditional search-based synthesis algorithms, since traditional algorithms are only designed to interpret demonstrations. They cannot handle NL descriptions that mention desired operations. Specifically, {\tool} encodes the demonstrations in NL and feeds them together with any NL description provided by users to the LLM for joint interpretation and synthesis. Finally, to help users effectively manage operations on multiple tables, {\tool} automatically tracks and visualizes the data provenance across multiple tables.


To achieve \textbf{G3}, an intuitive approach is to use LLMs to explain a synthesized script in natural language. However, LLMs are prone to hallucinations~\cite{leiser2024hill}. Besides, a recent study~\cite{gu2024analysts} shows that errors in LLM-generated code are often subtle and thus require users to carefully scrutinize each step of the analysis procedure. Inspired by recent systems that support step-by-step explanations~\cite{tian2024sqlucid, chen2023miwa}, we employ a grammar-based translation method to translate a generated script into step-by-step explanations in NL. This prevents LLM hallucination while providing a structured way for users to understand and verify the script. Another alternative approach is to render the program into a visual language~\cite{resnick2009scratch, chasins2018rousillon}. Although visually appealing, this approach still requires users to understand basic programming constructs such as variables and loops. In addition to NL explanations, we also allow users to run a synthesized script and observe its runtime behavior to confirm their understanding of the script and cross-validate its correctness. 

To achieve \textbf{G4}, we allow users to directly edit the NL explanation of the data wrangling script to fix any recognized errors or describe expected behavior. The updated explanation will be fed back to the LLM to refine the script. The rationale for this choice is two-fold. First, we want to continue leveraging the structured NL explanation in error correction, so users can immediately fix an error in situ as they are reading and scrutinizing each wrangling step in the explanation without context switching. Second, compared with two alternative approaches---providing more demonstrations or providing NL feedback in a conversation, editing individual steps in the explanation allows users to directly indicate where the errors are, which saves the effort for error localization and enables the LLM to perform a more focused script refinement. Nevertheless, we still support these two alternative error correction methods in {\tool} for flexibility. We measured the utility of each approach and reported the results in Section~\ref{sec:utility} and Figure~\ref{fig:logged_events}.



\section{SYSTEM DESIGN}
\begin{figure*}[ht!]
    \centering
    \includegraphics[width=\textwidth]{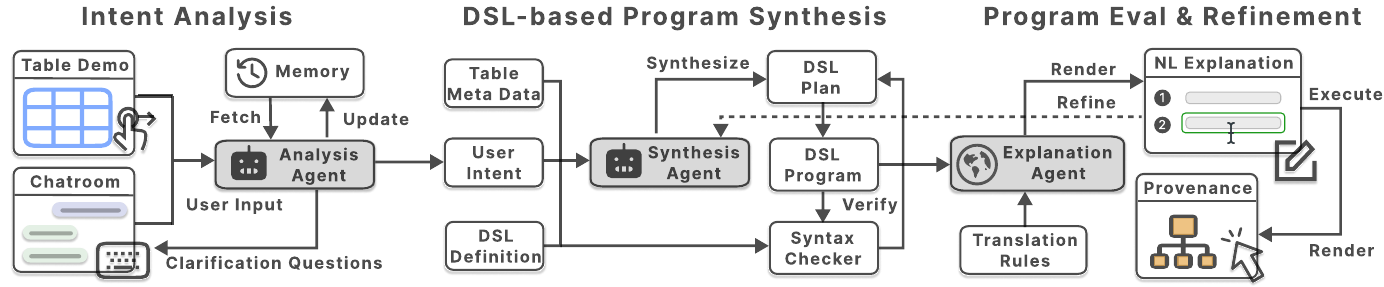}
    \captionsetup{font=small, labelfont=bf}
    \caption{Three core components: Intent Analysis, DSL-based Program Synthesis, and Program Evaluation \& Refinement.}
    \label{fig:system_arch}
\end{figure*}

As shown in Figure~\ref{fig:system_arch}, {\tool} includes three stages---\textit{Intent Analysis}, \textit{DSL-based Program Synthesis}, and \textit{Program Evaluation \& Refinement}. 
We adopt the multi-agent LLM framework to develop {\tool}. LLMs~\cite{achiam2023gpt, floridi2020gpt} have demonstrated \new{strong} capabilities in natural language understanding, contextual reasoning, and multi-turn conversations. For each stage, we develop a dedicated agent by leveraging the in-context learning capability of LLMs with few-shot prompting. Appendix~\ref{appendix:prompt} provides all prompt templates used in {\tool}.




\subsection{Intent Analysis}
\label{intent_analysis_implementation}

In this stage, {\tool} leverages an analysis agent to analyze user intent from the user's demonstrations and conversation history. If the intent is ambiguous, {\tool} generates clarification questions. After the user answers the CQs and there is no ambiguity, {\tool} generates a summary of user intent for the next stage to synthesize the desired data wrangling script.

\subsubsection{\textbf{User Demonstration}}

To capture user demonstrations, we set an event listener for every table entry in the uploaded tables. When a user modifies the table, the event listener logs the change in the demonstration history. 
For example, if a user changes a table entry from 0 to 3370 at row 5, column E, {\tool} logs this edit as \texttt{\small\{Row:5,Column:E,Old:0,New:3370,Type:Edit}\}. Specifically, {\tool} supports the following demonstrations on the uploaded tables:

\begin{enumerate}
\renewcommand{\labelenumi}{\arabic{enumi}.}
\item \textbf{\lbrack \textsc{INSERT}\rbrack}: insert empty columns or rows.
\item \textbf{\lbrack \textsc{DELETE}\rbrack}: delete columns or rows.
\item \textbf{\lbrack EDIT\rbrack}: edit specific cells.
\item \textbf{\lbrack COPY AND PASTE\rbrack}: copy and paste columns and rows.
\item \textbf{\lbrack DRAG AND DROP\rbrack}: drag and drop columns and rows.
\end{enumerate}

If an edit operation is applied to every cell in a row or column, {\tool} merges these individual operations to a single row/column edit operation in the demonstration history to save space.

\subsubsection{\textbf{Conversational Interface (Chatroom)}}
{\tool} provides a conversational interface that enables users to express their intent or feedback in NL. Moreover, when the user intent is unclear, {\tool} poses multiple-choice clarification questions to help users refine their intent. Each CQ includes different options and an ``Other (please specify)'' option, allowing users to provide customized responses when needed, as shown in Figure~\ref{fig:UI}\mycircle{f}.

\subsubsection{\textbf{Clarification Questions Generation \& Intent Summarization}}
\label{cq_generation}

{\tool} prompts an LLM to detect ambiguities and generate clarification questions. If no ambiguity is detected, the LLM summarizes the user intent into an NL sentence. Table~\ref{tab:intent_summarization_1} in Appendix~\ref{appendix:prompt} shows the prompt template. Specifically, user demonstrations are encoded as ``Table Diff'' in the INPUT section of the prompt, while NL inputs in the chatroom are encoded as ``User Instruction''. After the user answers a clarification question, the answer is appended to ``Chat History'' to identify any remaining ambiguities, as shown in Table~\ref{tab:intent_summarization_2}.

\subsection{DSL-based Program Synthesis}
\label{dsl_based_program_synthesis_implementation}
Once users define their intent, {\tool} employs a synthesis agent to generate step-by-step data wrangling scripts. 

\subsubsection{\textbf{Domain Specific Language}}
{\tool}'s Domain Specific Language is built on the foundational work of Potter's Wheel~\cite{raman2001potter, kandel2011wrangler}. We use this DSL because it has been adopted and validated in several existing work~\cite{kandel2011wrangler, jin2017foofah, guo2011proactive}. 
Since the original DSL only supports single-table operations, we extended the original DSL to support multi-table operations and advanced operations such as statistical tests. As shown in Appendix~\ref{appendix:dsl}, our extended DSL supports four types of operations: (1) \textit{Table-Level Operations}, which operate on entire tables (e.g., \texttt{\small merge}, \texttt{\small transpose}); (2) \textit{Column/Row-Level Operations}, which modify specific rows or columns (e.g., \texttt{\small insert}, \texttt{\small drop}); (3) \textit{Summarization Operations}, which aggregate data into new tables or statistical values (e.g., \texttt{\small aggregate}, \texttt{\small test}); and (4) \textit{String Operations}, which manipulate text transformations (e.g., \texttt{\small concatenate}, \texttt{\small format}). 

\subsubsection{\textbf{Plan Generation}} {\tool} adopts a self-planning strategy~\cite{jiang2024self, yao2022react} to first prompt the LLM to generate a synthesis plan for task decomposition. Each step in the plan contains a NL description of the desired action and the specific DSL operation that can achieve this action.
Table~\ref{prompt:plan} shows the prompt template.

\subsubsection{\textbf{DSL Program  Synthesis}}
After generating the step-by-step plan, {\tool} curates the arguments according to the DSL grammar specifications. To prevent potential hallucinations, we designed a syntax checker to statically check the correctness of the operation's arguments based on our DSL grammar. If verification fails, {\tool} appends error messages to the input and prompts the synthesis agent to generate a refined script. The prompt templates used in this step are detailed in Tables \ref{prompt:plan_with_error},~\ref{prompt:generate} and \ref{prompt:generate_dsl_with_error_message}.

\subsubsection{\textbf{DSL Program Execution}}
When users click{\raisebox{-0.9ex}{\includegraphics[height=1.2em]{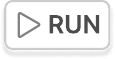}}}\space button to execute the script, {\tool} first imports the DSL function implementations into the execution environment. {\tool} then synthesizes a Python code snippet that uses these DSL functions, following the LLM prompt in Table~\ref{prompt:execute} in Appendix~\ref{appendix:prompt}. The snippet is executed using Python's \texttt{exec()} command~\cite{Python}. This approach facilitates dynamic code execution and handling of conditional operations. Note that the original table content will not be altered. Instead, new versions of the table are named with sequential version numbers (e.g., table\_v1, table\_v2). This idea is inspired by versioning systems~\cite{zhang2023vrgit, zolkifli2018version, kery2018story}.

\subsection{Program Evaluation \& Refinement}
\label{feedbacks_evaluation_implementation}
\subsubsection{\textbf{Step-by-Step Natural Language Explanation}}
\label{explanation_generation}
To help users understand the generated script, {\tool} translates each DSL statement into a natural language explanation. To prevent potential hallucinations, we design translation rules for each statement, as shown in Table \ref{table:dsl_translation} in Appendix~\ref{appendix:translation_rules}. For instance, given a DSL statement \texttt{delete\_table(sales.csv)} in a script, {\tool} translates the statement based on a text template—``\textit{Delete the table}'' + ``$X$''. In this template, the name $X$ will be replaced by the first argument \texttt{sales.csv} to compose the final string ``\textit{Delete the table \underline{sales.csv}}.'' For statements with conditions, {\tool} appends the condition string to the explanation. For instance, the final string will be ``\textit{Delete the table sales.csv \underline{if there are more than 30\% of missing values}.}''

\aptLtoX{\begin{table*}[h]
\begin{tabular}{|c|c|c|c|c| c |c|c|c|c|c|}
\hline
\textbf{StudentID} & \textbf{Name} & \textbf{Gender} & \textbf{Course A} & \textbf{Course B} && \textbf{StudentID} & \textbf{Name} & \textbf{Gender} & \textbf{Course C} & \textbf{Course D} \\
\hline
00336617 & John & Male & 9 & N/A && 00440033 & Alice & Female & 3 & 6 \\
\hline
\multicolumn{5}{|c|}{\ldots} && \multicolumn{5}{c|}{\ldots}\\
\hline
00770000 & Elisa & Female & 7 & 6 && 00993399 & Ben & Male & 5 & 5 \\
\hline
\multicolumn{11}{c}{$\downarrow$}\\
\end{tabular}
\begin{tabular}{|c|c|c|c|c| c |c|c|c|c|c|c|}
\hline
\multicolumn{2}{|c|}{\textbf{Name}} & \multicolumn{2}{|c|}{\textbf{StudentID}} & \multicolumn{2}{|c|}{\textbf{Course A}} & \multicolumn{2}{|c|}{\textbf{Course B}} & \multicolumn{2}{|c|}{\textbf{Course C}} & \multicolumn{2}{|c|}{\textbf{Course D}} \\
\hline
\multicolumn{2}{|c|}{Alice} & \multicolumn{2}{|c|}{00440033} & \multicolumn{2}{|c|}{5} & \multicolumn{2}{|c|}{7} & \multicolumn{2}{|c|}{3} & \multicolumn{2}{|c|}{6} \\
\hline
\multicolumn{12}{|c|}{\ldots} \\
\hline
\multicolumn{2}{|c|}{Yen} & \multicolumn{2}{|c|}{00990033} &\multicolumn{2}{|c|}{ 5} & \multicolumn{2}{|c|}{4} & \multicolumn{2}{|c|}{7} & \multicolumn{2}{|c|}{5} \\
\hline
\end{tabular}
\caption{The two uncleaned tables (shown at the top) and the desired cleaned table (shown at the bottom).}
\label{tab:course_ratings}
\end{table*}}{\begin{table*}[ht!]

\footnotesize
\setlength{\tabcolsep}{3pt}
\noindent
\begin{tabular}{|c|c|c|c|c|}
\hline
\textbf{StudentID} & \textbf{Name} & \textbf{Gender} & \textbf{Course A} & \textbf{Course B} \\
\hline
00336617 & John & Male & 9 & N/A \\
\hline
\multicolumn{5}{|c|}{\ldots} \\
\hline
00770000 & Elisa & Female & 7 & 6 \\
\hline
\end{tabular}
\hspace{0.2cm}
\begin{tabular}{|c|c|c|c|c|}
\hline
\textbf{StudentID} & \textbf{Name} & \textbf{Gender} & \textbf{Course C} & \textbf{Course D} \\
\hline
00440033 & Alice & Female & 3 & 6 \\
\hline
\multicolumn{5}{|c|}{\ldots} \\
\hline
00993399 & Ben & Male & 5 & 5 \\
\hline
\end{tabular}

\vspace{0.1cm}  
$\downarrow$

\vspace{0.1cm}  
\begin{tabular}{|c|c|c|c|c|c|}
\hline
\textbf{Name} & \textbf{StudentID} & \textbf{Course A} & \textbf{Course B} & \textbf{Course C} & \textbf{Course D} \\
\hline
Alice & 00440033 & 5 & 7 & 3 & 6 \\
\hline
\multicolumn{6}{|c|}{\ldots} \\
\hline
Yen & 00990033 & 5 & 4 & 7 & 5 \\
\hline
\end{tabular}
\captionsetup{font=small}
\caption{The two raw tables (shown at the top) and the desired cleaned table (shown at the bottom).}
\label{tab:course_ratings}
\end{table*}}


\subsubsection{\textbf{DSL Program Refinement}}
Due to potential hallucinations of the LLM agents used in our system, {\tool} might synthesize erroneous scripts that are not desired. To help users refine the synthesized program, {\tool} allows users to refine the synthesized scripts through the following actions (Figure 3\mycircle{g}).

\begin{enumerate}
\renewcommand{\labelenumi}{\arabic{enumi}.}
\item \textbf{\lbrack \textsc{EDIT}\rbrack}: edit a step in the NL explanation.
\item \textbf{\lbrack \textsc{DELETE}\rbrack}: delete a step.
\item \textbf{\lbrack ADD\rbrack}: add a new step in NL.
\item \textbf{\lbrack SAVE\rbrack}: save the current script.
\item \textbf{\lbrack REMOVE\rbrack}: remove the current script.
\item \textbf{\lbrack REGENERATE\rbrack}: regenerate the script.
\end{enumerate} 
If the user only adds a step without changing other steps, {\tool} will prompt the LLM to simply update the script with the newly added step using the prompt template in Table~\ref{prompt:update_dsl}. For other types of edits, {\tool} will prompt the LLM to regenerate the DSL script based on the previous script using the prompt template in Table \ref{prompt:edit_dsl}. In both cases, {\tool} will prompt the LLM to re-summarize the user intent for further usage, e.g., in the next round of script generation or refinement.


\subsubsection{\textbf{Data Provenance View}}

Multi-table operations usually involve complex data dependencies. This may increase the user’s cognitive overload to memorize each change and the inter-dependency between tables. Thus, {\tool} visualizes \textit{data provenance} to allow users to track different versions of tables and their dependencies. When a user executes the scripts, {\tool}'s backend calculates the data dependency by analyzing the data flow between different tables. For instance, when \textsc{C = merge(A, B)}, {\tool} will render two directional edges from $\textsc{A} \rightarrow \textsc{C}$ and from $\textsc{B} \rightarrow \textsc{C}$. Each node represents a specific version of a table, with directional edges showing dependencies between tables. The dependency is done at the table level, not at a more fine-grained level such as row or column. This is because finer granularity would make the provenance view large and overwhelming, also known as \textit{flood of information}~\cite{ahlberg1994visual, borgman1986online}. To see the content of a certain table at a certain version, users can click nodes in the \textit{Data Provenance} view panel. The corresponding table content will appear in the \textit{Table View} panel. This creates a visual correspondence, or \textit{brushing}~\cite{becker1987brushing},  allowing users to interactively check corresponding table content.

\subsection{System Implementation}
{\tool}'s frontend is developed with React and Vite~\cite{Vite}. For state management, {\tool} employs Zustand~\cite{Zustand}. {\tool} uses Tailwind CSS~\cite{TailwindCSS} to create a responsive user interface. The user interface incorporates Handsontable~\cite{Handsontable}, providing an intuitive experience for spreadsheet tables, as well as React Flow~\cite{ReactFlow}, providing Node-Based UIs for the data provenance feature. {\tool}'s backend incorporates FastAPI~\cite{FastAPI}, a web framework for building backend APIs with Python. We use OpenAI's gpt-4o-mini~\cite{hurst2024gpt} as the LLM for {\tool}. We set the \texttt{temperature} to $0$ to make our experiments reproducible while keeping all other hyperparameters at their default settings. In particular, the default setting of \texttt{top\_p} is $1$, \texttt{frequency\_penalty} is $0$, and \texttt{presence\_penalty} is $0$.



\begin{figure*}[ht!]

    \centering
    \includegraphics[width=\textwidth]{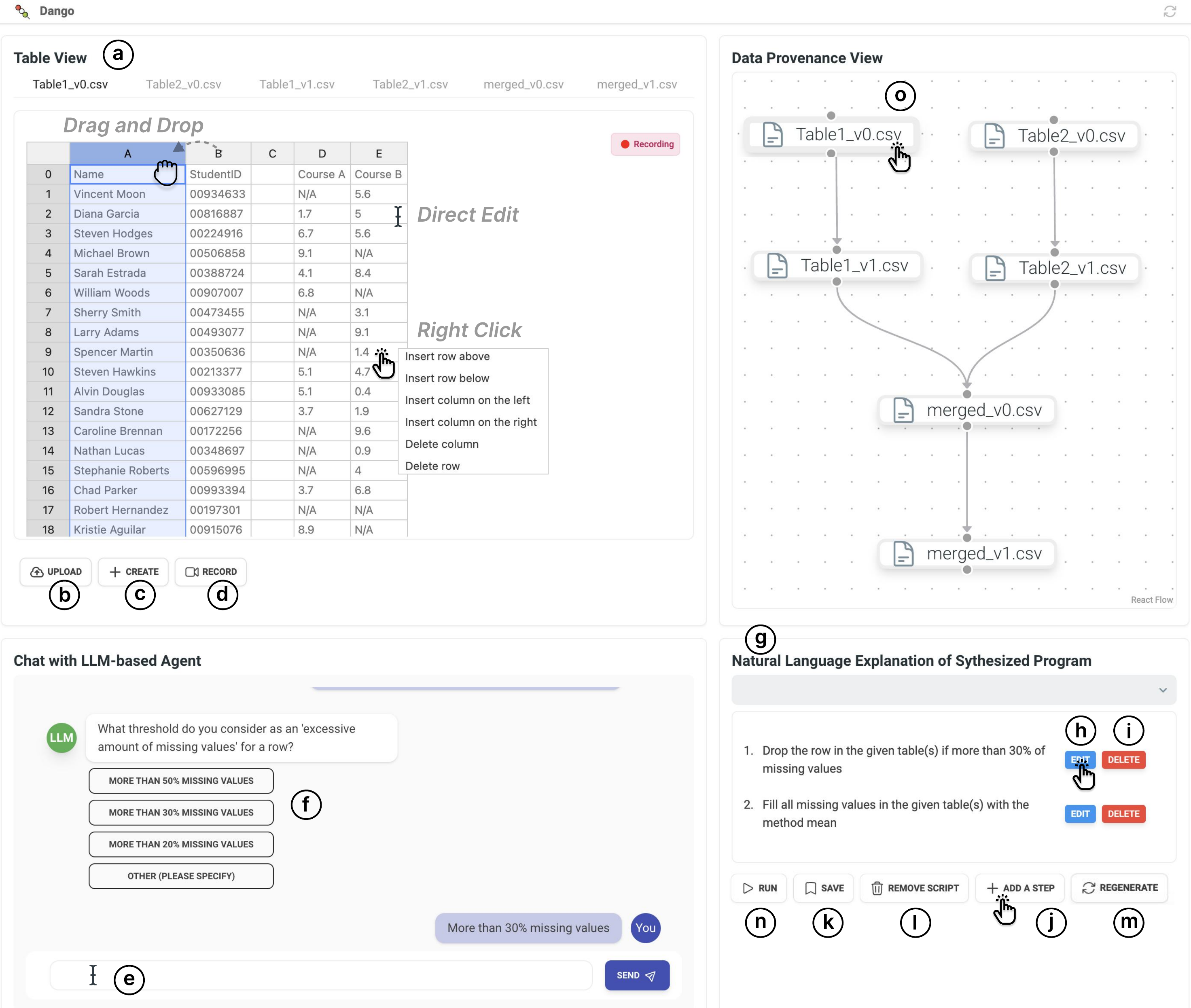}
    \captionsetup{font=small, labelfont=bf}
    \caption[]{User interface of {\tool}. In the table view (\mycircle{a}),  users can upload tables (\mycircle{b}) or create new tables (\mycircle{c}). Then, they can click the {\raisebox{-0.9ex}{\includegraphics[height=1.2em]{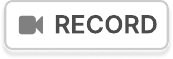}}}\space button (\mycircle{d}) and start demonstrating their desired actions. Alternatively, they can express desired actions in natural language in a chatbox (\mycircle{e}).  {\tool} will interpret the demonstrations and/or NL descriptions in the backend and generate multiple-choice clarification questions when needed (\mycircle{f}). Furthermore, to help users understand and validate the synthesized script, {\tool} explains it in NL step by step  (\mycircle{g}). Users can directly edit a step in natural language (\mycircle{h}), delete a step (\mycircle{i}), add a new step  (\mycircle{j}), save the script (\mycircle{k}), remove the script (\mycircle{l}), or regenerate the script (\mycircle{m}). Users can click the {\raisebox{-0.9ex}{\includegraphics[height=1.2em]{figures/runbtn.pdf}}}\space button (\mycircle{n}) to execute the script on copies of the original tables and verify its behavior without messing up the original demonstrations. {\tool} also renders a data provenance view to track the transformations performed on each table (\mycircle{o}). Users can click table nodes, and the corresponding table content will appear in the table view.}
    
    \label{fig:UI}

\end{figure*}

\begin{figure*}[ht!]

    \centering
    \includegraphics[width=\textwidth]{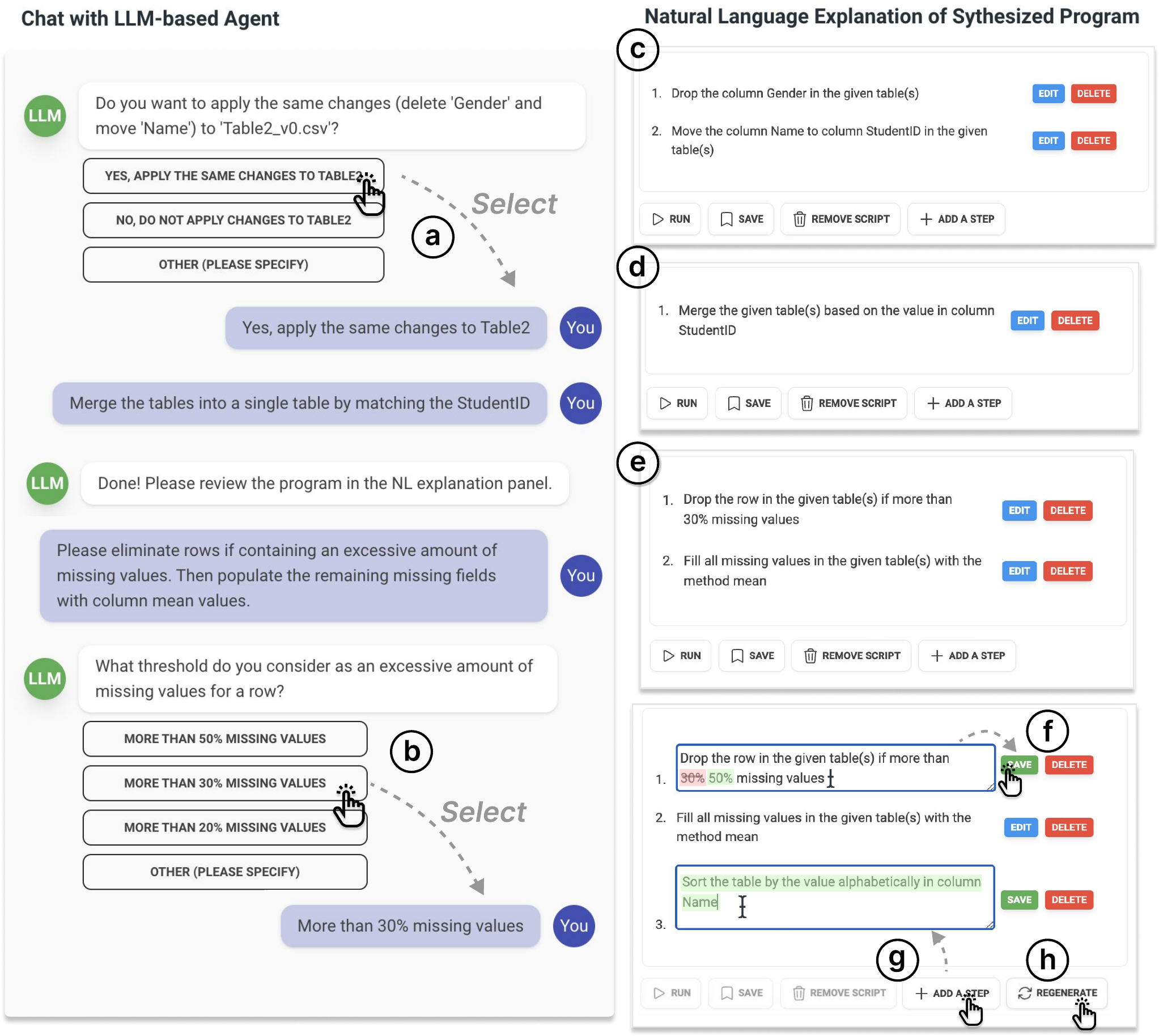}
    \captionsetup{font=small, labelfont=bf}
    \caption{This figure shows a usage scenario of users using a chatroom to clarify their intent using NL prompts and answering multiple-choice clarification questions (left-hand side). Users can easily understand the program behavior by reading the step-by-step NL explanations. They can refine their program by directly editing the step-by-step NL explanations (right-hand side)}
    
    \label{fig:usage}

\end{figure*}

\section{USAGE SCENARIO}

Beth is a professor in the Department of Education who wants to study the students' ratings on different courses to improve teaching quality and student satisfaction. She has run two separate surveys for different courses and collected student ratings. The raw data have been downloaded into two separate tables, as shown in Table~\ref{tab:course_ratings}. Since manually cleaning data is time-consuming, she decides to use {\tool} to generate an automated script.

She first uploads two raw tables to {\tool}. Next, she clicks the {\raisebox{-0.9ex}{\includegraphics[height=1.2em]{figures/recordbtn.pdf}}\space button (Figure~\ref{fig:UI}\mycircle{d}) and demonstrates to {\tool} how to clean the data. She deletes the {\small\textsf{Gender}}} column and then moves the {\small\textsf{Name}} column ahead of {\small\textsf{StudentID}}. After her demonstration, {\tool} generates a multiple-choice clarification question in the chatroom, asking if she wants to apply the same changes to Table 2. Beth confirms by selecting the ``Yes'' option (Figure~\ref{fig:usage}\mycircle{a}). {\tool} then generates the DSL script and renders the step-by-step NL explanation of the script, as shown in Figure~\ref{fig:usage}\mycircle{c}.

Beth reads this step-by-step NL explanation and confirms that it matches her intent. She clicks the {\raisebox{-0.9ex}{\includegraphics[height=1.2em]{figures/runbtn.pdf}}}\space button and soon notices that the data provenance view is rendered (Figure~\ref{fig:UI}\mycircle{o}). She clicks the ``Table2\_v1.csv'' node in the data provenance view and confirms that the transformations to Table 2 are correct. 

Beth now wants to merge these two tables. She types in the chatroom: ``\textit{Merge the tables into a single table by matching the StudentID}''. {\tool} then generates a script and its explanation  (Figure~\ref{fig:usage}\mycircle{d}). This looks correct to Beth, so she clicks {\raisebox{-0.9ex}{\includegraphics[height=1.2em]{figures/runbtn.pdf}}}\space to perform the action.
Then, Beth notices some missing values in the merged table. She then types,  ``\textit{Please eliminate rows if containing an excessive amount of missing values. Then populate the remaining missing fields with column mean values.}'' {\tool} then generates a clarification question to ask Beth to specify the threshold for excessive missing values (Figure~\ref{fig:usage}\mycircle{b}). Beth selects 30\% as the threshold. {\tool} generates the new script accordingly based on the clarification (Figure~\ref{fig:usage}\mycircle{e}).

Later, Beth changes her mind and clicks the  {\raisebox{-0.9ex}{\includegraphics[height=1.2em]{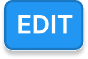}}}\space button (Figure~\ref{fig:usage}\mycircle{f}), adjusting the threshold from 30\% to 50\%. She then clicks the  {\raisebox{-0.9ex}{\includegraphics[height=1.2em]{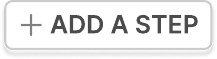}}}\space button (Figure~\ref{fig:usage}\mycircle{g}) and types in a new step: ``\textit{Sort the table alphabetically by the values in the column Name.}'' Next, she clicks the {\raisebox{-0.9ex}{\includegraphics[height=1.2em]{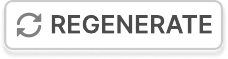}}}\space button (Figure~\ref{fig:usage}\mycircle{h}) to obtain the new script. Lastly, Beth runs the script and gets the final table.



\section{EVALUATION}
We investigated five research questions in the evaluation of {\tool}. To evaluate the usability and efficiency of \tool, we conducted an IRB-approved within-subjects user study with 38 participants on 7 data wrangling tasks (\textbf{RQ1-4}). To understand the generalizability of {\tool}, we conducted a quantitative study with 24 additional tasks (\textbf{RQ5}).

\begin{itemize}[leftmargin=10pt]

\item \textbf{RQ1.} \textit{How different interaction paradigms of {\tool} can help users in data wrangling tasks?} \new{{\tool} supports multiple interaction paradigms for user intent clarification and refinement. This RQ aims to explore how these interaction paradigms contribute to data wrangling effectiveness by comparing {\tool} with its variants.} 

\item \textbf{RQ2.} \textit{How do users perceive and value different features of {\tool}?} \new{This RQ evaluates user perceptions of different features in {\tool} by analyzing feature ratings and qualitative feedback.}

\item \textbf{RQ3.} \textit{Why does {\tool} help users perform data wrangling tasks more effectively?}  \new{This RQ aims to get a deeper understanding of {\tool}'s effectiveness by analyzing and summarizing the common reasons why participants liked {\tool} mentioned in the post-study survey.}

\item \textbf{RQ4.} \textit{How does user expertise level influence task performance?} \new{We envision that {\tool} will not only be helpful for end-users but also improve the productivity of experienced users. Thus, we recruited participants with different levels of expertise in data wrangling in our user study. This RQ examines whether user expertise affects user performance.}
\item \textbf{RQ5.} \textit{How well does {\tool} generalize to a broader range of data wrangling tasks?} \new{To evaluate the generalizability of {\tool}, we conducted a systematic evaluation across an additional set of 24 data wrangling tasks. Furthermore, we analyze the distribution of generated DSL statements and the effectiveness of clarifying questions (CQs) in this broader task set.}
\end{itemize}

\subsection{User Study Conditions}
\label{sec:conditions}
We designed three user study conditions, each of which represents a specific design of {\tool}. The comparison of user performance across these conditions helped us understand the effectiveness of different interaction paradigms for users to clarify their intent and provide feedback to refine the generated script. The data provenance view is enabled in all three conditions.  

\noindent\textbf{Condition A (Demonstration + Conversational Interface + NL summary):} This condition represents a naive design of {\tool}. It simply integrates the two interaction modalities and prompts the LLM to generate the NL summary of a synthesized script. Table~\ref{prompt:dsl_to_nl} shows the promote template to generate the NL summary. This NL summary does not follow the step-by-step structure and there is no special treatment to mitigate LLM hallucination. To clarify user intent or refine the synthesized script, users can provide new demonstrations and provide feedback in the chatroom. However, they cannot edit the NL summary to provide feedback. 

\noindent\textbf{Condition B (Demonstration + Conversational Interface + Step-by-Step NL):} This condition represents an enhanced design of the NL feedback mechanisms in {\tool}. In this condition, {\tool} generates the step-by-step NL explanation of  a script using the rule-based method proposed in Section~\ref{explanation_generation}. This design enables a more faithful explanation with no LLM hallucinations compared to the NL summary in Condition A. Furthermore, users can provide more fine-grained feedback to refine the synthesized script by directly editing specific steps in the NL explanation.

\noindent\textbf{Condition C (Demonstration + Conversation Interface+ Step-by-Step NL + CQs):} This condition represents the final design of {\tool}. Compared with Condition B, this condition allows {\tool} to proactively ask clarification questions to resolve ambiguities, as proposed in Section~\ref{cq_generation}. This transforms from user-initiative interaction in Condition B to a collaborative mixed-initiative approach, where the system and user work together to clarify and refine the intent.

\subsection{User Study Participants}
We recruited a total of 38 participants (11 female, 27 male) through a mailing list at Purdue University. 32\% of participants had no programming experience at all or had only taken an introductory programming course for one semester. 37\% had one or more years of programming experience but less than 5 years. 31\% had five or more years of programming experience. 42\% of participants were undergraduate students, 11\% were master's students, and 47\% were PhD students. These students were from diverse majors, including Computer Science, Mechanical Engineering, Industrial Engineering, Materials Engineering, Statistics, Chemical Engineering, Physics, Agriculture, Business, Hospitality, Communication, and Game Design. Table~\ref{tab:demographics} in Appendix~\ref{appendix:userstudy} shows the detailed demographic information of each participant.

\subsection{User Study Tasks}
To develop the user study tasks, we first identified 20 data wrangling tasks from previous data wrangling work ~\cite{kandel2011wrangler, li2024sheetcopilot, chen2022rigel} and extended our tasks with 21 multi-table tasks based on Kasica et al.'s recent empirical work~\cite{kasica2020table}. Then we discussed and refined these tasks through three meetings with a domain expert from the College of Education at Purdue University, who regularly performs data wrangling tasks such as cleaning student survey responses in spreadsheets. During the discussions, we removed 13 tasks that the domain expert thought were too simple or uncommon in practice. The domain expert also suggested 3 new tasks with conditional operations and statistical tests. From a total of 31 candidate tasks, we selected 7 most representative data wrangling tasks based on the suggestion of the domain expert, including 2 single-table tasks and 5 multi-table tasks. Among them, 2 tasks involve conditional operations, and 1 task focuses on statistical testing Table~\ref{tab:data-tasks} in Appendix~\ref{appendix:userstudy} provides a detailed description of each task.

\subsection{User Study Procedure}
Each participant came to our lab for a reserved one-hour session. Participants then signed the consent form. To reduce demand characteristics~\cite{orne2017social}, we told participants that our goal was to understand the influence of three different conditions of an existing tool. We did not tell them that we developed this tool or which condition represented the final design. Furthermore, to mitigate the learning effect~\cite{lazar2017research}, both task and condition assignment orders were counterbalanced across participants. As compensation, each participant received a \$25 Amazon gift card.

\vspace{1mm}
\noindent\textbf{\textit{Timed Data Wrangling Tasks.}}
 Participants watched a 5-minute tutorial video and spent about 5 minutes becoming familiar with the tool. Then they completed 3 randomly assigned data wrangling tasks selected from 7 user study tasks with 3 different conditions. These assignments were counterbalanced across 3 conditions and 7 tasks, resulting in 5 or 6 trials per task per condition. Since only tasks 2 and 6 are single-table tasks, we also counterbalanced the task assignments so that each user experienced at least two multi-table tasks. Participants will first read the task description and make sure they understand the tasks. We explicitly told participants not to directly copy and paste the task description to the chatroom. A task was considered failed if participants did not generate a script that could correctly clean the data within 10 minutes. 

 \vspace{1mm}
\noindent\textbf{\textit{Post-task Surveys.}}
After completing each task, participants filled out a post-task survey to give feedback, as described in Table~\ref{tab:posttask} in Appendix~\ref{appendix:userstudy}. The survey first asked users to report their confidence in the generated data wrangling scripts on a 7-point scale (1---very low confidence, 7---very high confidence). Then it asked users to rate the usefulness of key features in each condition on a 7-point scale. It also asked users what they liked or disliked about the tool and what they wished they had. To evaluate the cognitive load of using a tool, we included five NASA Task Load Index questions \cite{hart1988development} as part of the post-task survey.

\vspace{1mm}
\noindent\textbf{\textit{Post-study Survey.}}
After completing all tasks, participants completed a post-study survey comparing {\tool} across conditions. The survey included questions on which condition was most helpful, reasons for their preferences, and open-ended feedback. Table~\ref{tab:poststudy} in Appendix~\ref{appendix:userstudy} lists the questions from the survey.

\subsection{User Study Measurements \& Data Analysis}
\label{sec:analysis}

We collected both quantitative and qualitative data from the user study. We measured task completion time and the number of attempts per task. We considered a participant made another attempt when they (1) submitted an incorrect script to the experimenter or (2) started over from scratch. To gain a deeper understanding of user behavior patterns in different conditions, we measured the utility of each key feature by analyzing user events in system logs, such as starting a new demonstration and editing the NL explanation. 

Moreover, {\tool} heavily relies on LLMs to analyze user intent and generate clarification questions and data wrangling scripts. Since LLMs may misinterpret user intent and hallucinate, it is important to understand how often they hallucinate during a task session. Thus, the second author watched all video recordings and examined the LLM-generated clarification questions and data wrangling scripts. A hallucination is detected if a clarification question is irrelevant to user intent or if the generated script is incorrect.   

To analyze the quantitative ratings collected from the post-task surveys, we used ANOVA tests~\cite{fisher1966design} to examine the statistical significance of their mean differences. To analyze the open-ended feedback, the first author conducted open coding~\cite{vaismoradi2013content} to identify themes in participants' responses, followed by thematic analysis~\cite{braun2006using}. \new{The last author reviewed the coding results and discussed with the first author to refine the initial codes and themes. The first author then incorporated feedback, made revisions and adjustments, and finalized the themes before writing the reports.}


\begin{table*}[ht!]
\centering
\setlength{\tabcolsep}{4pt} 
\renewcommand{\arraystretch}{1.4} 
\scriptsize 
\begin{tabular}{|l|*{21}{c|}}
\hline
\multirow{3}{*}{\textbf{Measure}} & \multicolumn{3}{c|}{\textbf{Task 1}} & \multicolumn{3}{c|}{\textbf{Task 2}} & \multicolumn{3}{c|}{\textbf{Task 3}} & \multicolumn{3}{c|}{\textbf{Task 4}} & \multicolumn{3}{c|}{\textbf{Task 5}} & \multicolumn{3}{c|}{\textbf{Task 6}} & \multicolumn{3}{c|}{\textbf{Task 7}} \\
\cline{2-22}
 & \multicolumn{3}{c|}{\textbf{Condition}} & \multicolumn{3}{c|}{\textbf{Condition}} & \multicolumn{3}{c|}{\textbf{Condition}} & \multicolumn{3}{c|}{\textbf{Condition}} & \multicolumn{3}{c|}{\textbf{Condition}} & \multicolumn{3}{c|}{\textbf{Condition}} & \multicolumn{3}{c|}{\textbf{Condition}} \\
\cline{2-22}
 & \textbf{A} & \textbf{B} & \textbf{C} & \textbf{A} & \textbf{B} & \textbf{C} & \textbf{A} & \textbf{B} & \textbf{C} & \textbf{A} & \textbf{B} & \textbf{C} & \textbf{A} & \textbf{B} & \textbf{C} & \textbf{A} & \textbf{B} & \textbf{C} & \textbf{A} & \textbf{B} & \textbf{C} \\
\hline

Time & 8:56 & 4:29 & 5:18 & 3:57 & 3:20 & 1:34 & 5:42 & 5:46 & 3:40 & 4:40 & 3:06 & 3:31 & 7:46 & 6:49 & 4:29 & 5:10 & 5:30 & 2:40 & 7:42 & 6:35 & 3:33 \\
\hline
Attempts & 2.67 & 1.60 & 2.00 & 1.00 & 1.50 & 1.00 & 1.75 & 1.75 & 1.67 & 1.83 & 1.50 & 1.00 & 2.00 & 2.00 & 1.75 & 2.00 & 1.80 & 1.33 & 2.40 & 1.40 & 1.20 \\
\hline
\# of CQs & -- & -- & 0.40 & -- & -- & 0.60 & -- & -- & 1.33 & -- & -- & 0.50 & -- & -- & 3.00 & -- & -- & 1.00 & -- & -- & 1.40 \\
\hline
\end{tabular}
\captionsetup{font=small}
\caption{The average task completion time, average number of attempts, and average number of CQ generated.}
\label{tab:performance}
\end{table*}




\section{EVALUATION RESULTS}
 \subsection{RQ1: Effectiveness of Different Interaction Paradigms}
\label{sec:performance}
\subsubsection{
{\bf \textit{Overall Performance.}}}
All participants successfully completed the tasks in Conditions B and C. However, in Condition A, two participants were unable to generate the correct scripts within the 10-minute time limit. The average completion time using {\tool} in Condition C is 3 min 30 sec, representing a 45\% decrease compared to condition A and a 32\% decrease compared to condition B. An ANOVA test shows that the mean differences in time used are statistically significant ($p$-value = 1.76e-05). When using {\tool} in Condition C, participants made an average of 1.43 attempts across all tasks, compared to 2 and 1.66 attempts in Conditions A and B, respectively. An ANOVA test shows that the mean differences in number of attempts are statistically significant ($p$-value = 2.89e-02).

The average percentage of hallucination in each task session is 0.65 in Condition A, 0.59 in Condition B, and 0.18 in Condition C. An ANOVA test revealed that the mean differences of the hallucination rate across these three conditions are statistically significant ($p$-value = 1.32e-02). Furthermore, Figure~\ref{fig:logged_events} shows the relative temporal distribution of hallucinated scripts in each task session in each condition. These results demonstrate the effectiveness of proactively asking clarification questions in reducing hallucinations. 

\subsubsection{{\bf \textit{Task-Specific Performance.}}} Table \ref{tab:performance} shows user performance in different tasks and conditions. For Tasks 2, 5, 6, and 7, Condition C significantly reduced the task completion time compared to Condition A. Pairwise ANOVA tests show that the mean
differences are statistically significant ($p$-value=4.89e-02, $p$-value=4.22e-02, $p$-value=2.13e-04, $p$-value=2.77e-03, respectively). For Tasks 5, 6, and 7, Condition C significantly reduced the task completion time compared to Condition B. Pairwise ANOVA tests show that the mean differences are statistically significant ($p$-value=4.82e-02, $p$-value=1.42e-02, $p$-value=3.46e-02, respectively). There are no statistically significant differences in Tasks 1, 3, and 4 since these tasks are easier compared with other tasks. Tasks 5, 6, and 7 include complex operations or conditional operations. This suggests that condition C's advantages become more prominent in complex scenarios that require users to clarify their intent.

\subsubsection{ {\bf \textit{User Confidence \& User Preference.}}}  As shown in Figure \ref{fig:confidence_ratings_chart}, participants expressed a higher level of confidence when using {\tool} in Condition C  (\textit{Mean}: 5.31~vs.~6.18 vs.~6.34). The mean differences are statistically significant (One-way ANOVA: $p$-value= 1.64e-04). When asked which condition they preferred for data wrangling, 82\% of participants chose Condition C.

\aptLtoX[graphic=no,type=html]{\begin{figure}[h!]
    \centering
    \includegraphics[width=\textwidth]{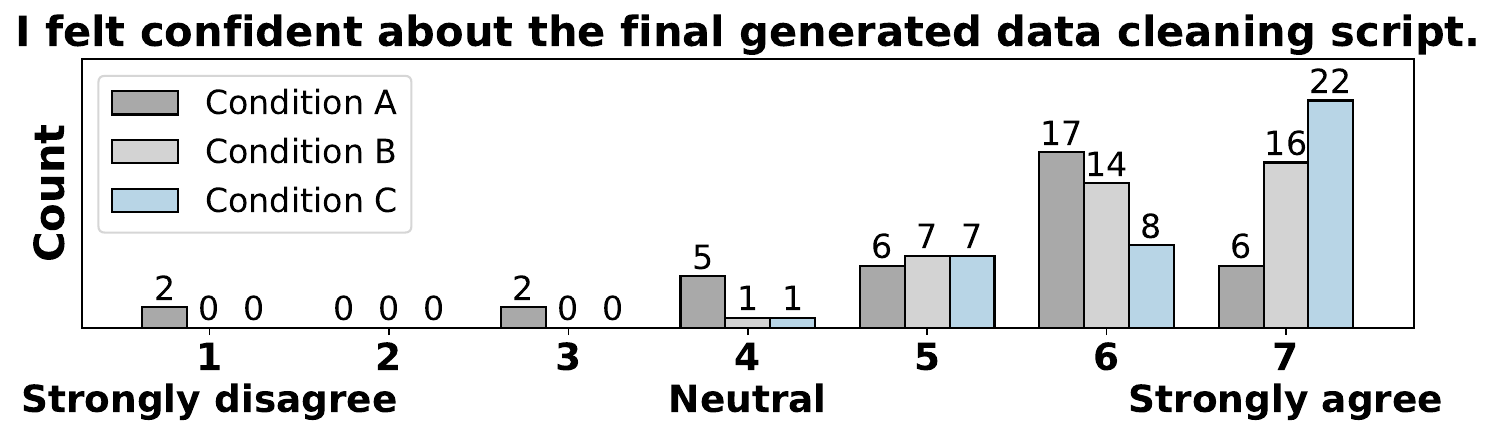}
    \captionsetup{font=small}
    \caption{User confidence on the data wrangling scripts when using conditions A, B, and C.}
    \label{fig:confidence_ratings_chart}
  \end{figure}
  \hfill
  \begin{figure}
    \centering
    \includegraphics[width=\textwidth]{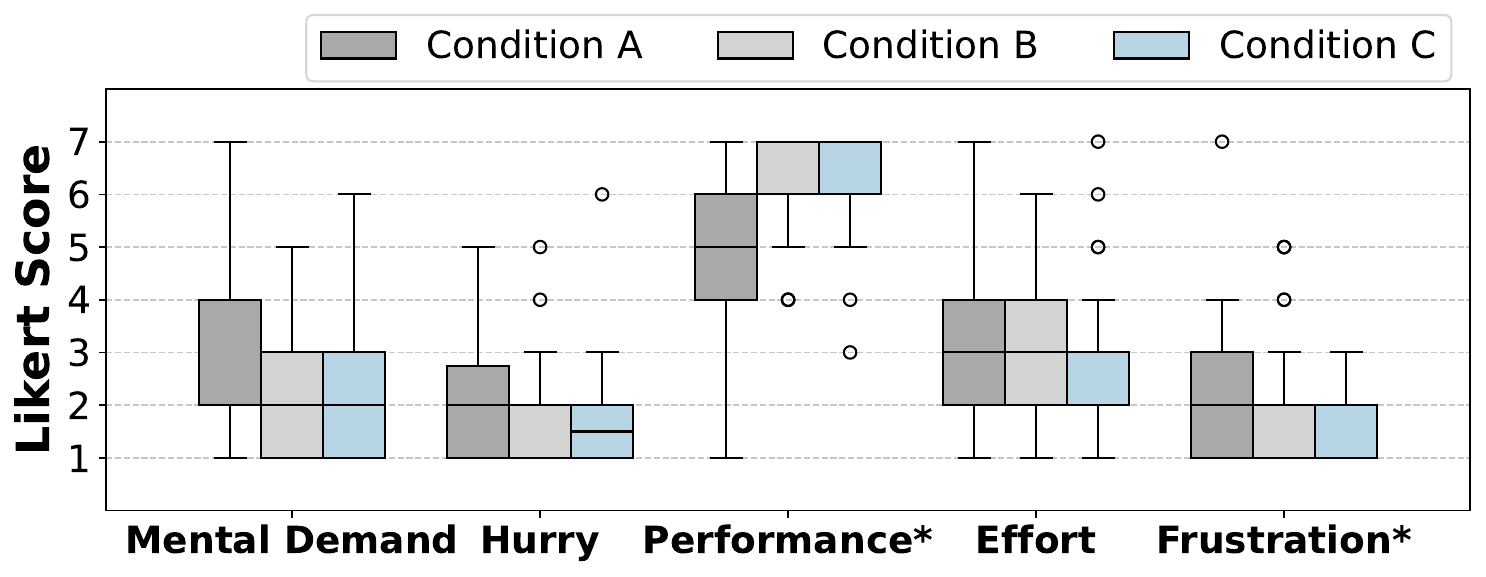}
    \captionsetup{font=small}
    \caption{User responses to the NASA TLX questionnaire (*: $p$-value < 0.05 based on ANOVA test).}
    \label{fig:TLX}
\end{figure}}{\begin{figure}[h!]
  \centering
  \begin{minipage}{0.48\textwidth}
    \centering
    \includegraphics[width=\textwidth]{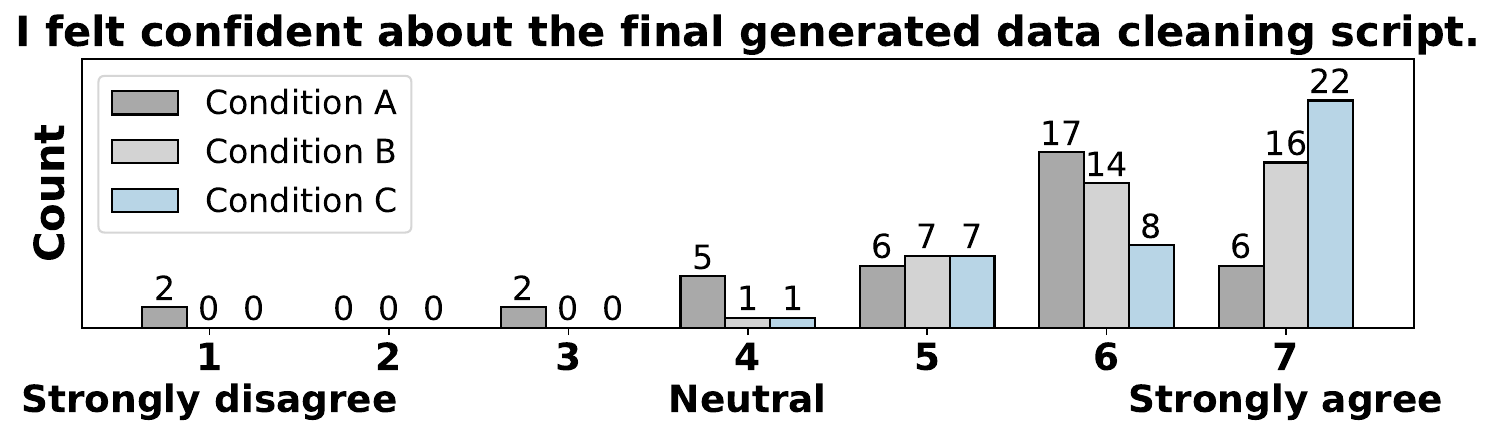}
    \captionsetup{font=small}
    \caption{User confidence on the data wrangling scripts when using conditions A, B, and C.}
    \label{fig:confidence_ratings_chart}
  \end{minipage}
  \hfill
  \begin{minipage}{0.48\textwidth}
    \centering
    \includegraphics[width=\textwidth]{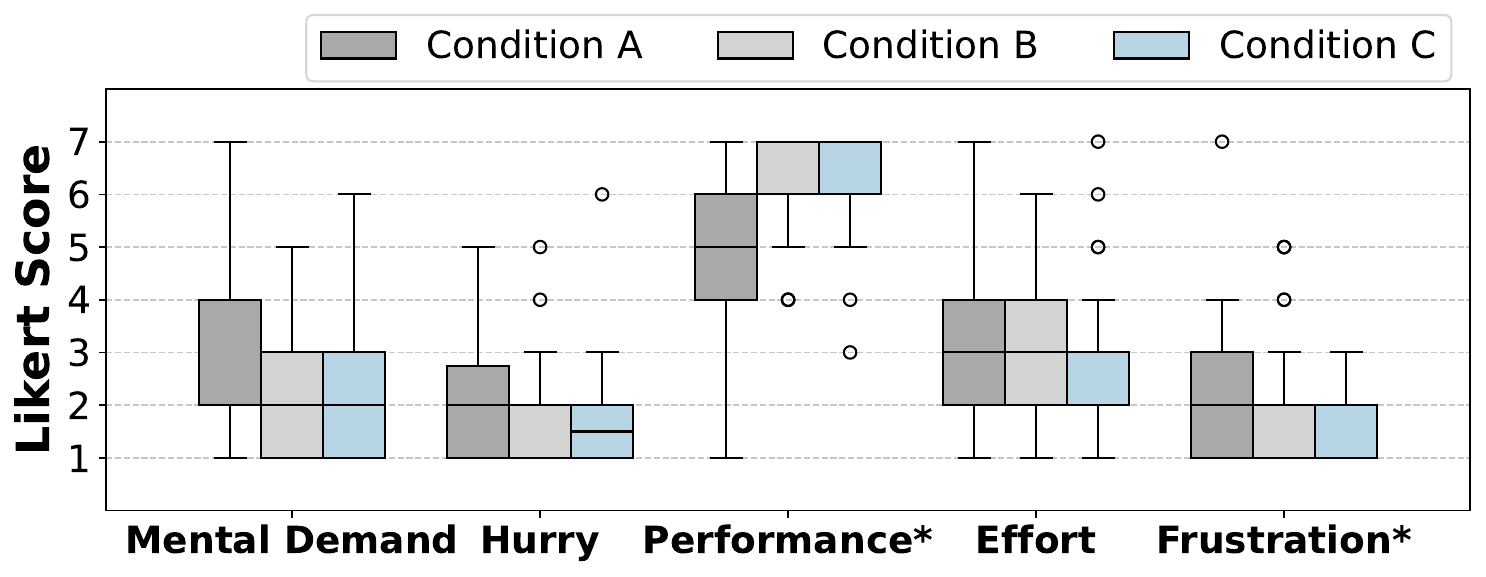}
    \captionsetup{font=small}
    \caption{User responses to the NASA TLX questionnaire (*: $p$-value < 0.05 based on ANOVA test).}
    \label{fig:TLX}
  \end{minipage}
\end{figure}}

\subsubsection{ {\bf \textit{Cognitive Overhead.}}} As shown in Figure \ref{fig:TLX}, we detected significant differences in \textit{Performance} and \textit{Frustration}. The ANOVA tests reveal that the mean differences are statistically significant for both metrics ($p$-value = 1.42e-05, and 2.98e-02, respectively). For \textit{Mental Demand}, \textit{Hurry}, and \textit{Effort}, we did not detect significant differences ($p$-value = 1.3e-01, 3.71e-01, and 2.77e-01, respectively). 

\begin{figure*}[t!]

    \centering
    \includegraphics[width=\textwidth]{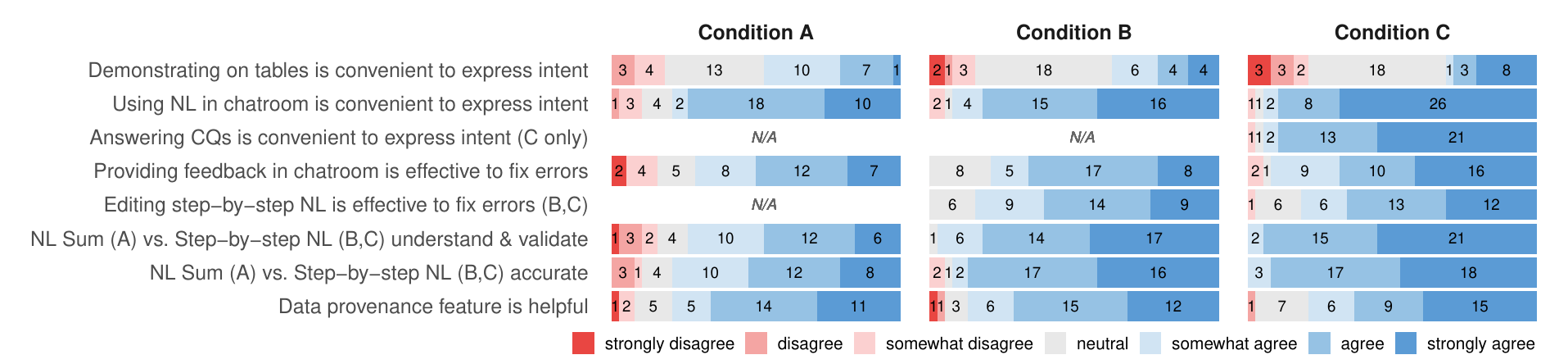}
    \captionsetup{font=small, labelfont=bf}
    \caption{7-point Likert scale evaluations (1---strongly disagree, 7---strongly agree) of user-perceived convenience, and effectiveness of {\tool}’s key features across three conditions.}    
    
    \label{fig:likert_ratings}

\end{figure*}

\subsection{RQ2: User Perceptions}

In the post-task survey, participants rated the usefulness of key
features of {\tool} in 7-point Likert scale questions (1---strongly disagree, 7---strongly agree), as shown in Figure~\ref{fig:likert_ratings}.

\subsubsection{ {\bf \textit{Participants' perception on different intent expression features}}}
{\tool} supports three features that help users express their intent. First, users can perform demonstrations on uploaded tables. Second, users can use the chatroom to prompt their intent in natural language. Third, if their initial input is ambiguous, they can clarify their intent by answering multiple-choice clarification questions posed by the {\tool}.

\vspace{1mm}
\noindent{\bf \textit{Demonstrations.}} Participants expressed mixed opinions about demonstrating on the uploaded tables, as shown in Figure~\ref{fig:likert_ratings}.
Only 21\% of participants in Condition A, 21\% in Condition B, and 29\% in Condition C agreed or strongly agreed that demonstration was a convenient way to express intent. Participants mentioned that  \new{demonstrating on tables was fast and straightforward for simpler tasks (P5), but error-prone for complex tasks if demonstrations are not clear enough (P1).}  This suggests demonstrations might be more effective for simpler tasks but are error-prone for complex tasks.

\vspace{1mm}
\noindent{\bf\textit{Using NL Prompts in the Chatroom.}} Compared to demonstrations, significantly more users preferred to express their intent in NL. 74\% of participants in Condition A, 82\% in Condition B, and 89\% in Condition C agreed or strongly agreed that using NL in the chatroom was convenient to express their intent. Participants appreciated the chatroom feature for \new{allowing easy expression of requests in natural language (P16), facilitating straightforward communication (P3), and enabling clear articulation of data cleaning actions (P27).} The result suggests that the conversational UI is effective in communicating user intent. 

\vspace{1mm}
\noindent{\bf \textit{Answering CQs.}}
As shown in 
Figure~\ref{fig:likert_ratings}, 89\% of the participants in Condition C agreed or strongly agreed that answering CQs is convenient to express their intent. Participants mentioned that \new{it is helpful to have {\tool} first ask clarification questions because it shows its understanding of their intent (P11) and reduces the burden to type more detailed clarification instructions when the output is not desired (P12, P22).} This result suggests that proactively asking clarification questions can effectively help users clarify intent and complement NL prompting.

\subsubsection{ {\bf \textit{Participants' Perceptions of Different Error-Fixing Features}}} {\tool} supports two features that help users fix their programming errors. First, in all conditions, users can use NL prompts to provide feedback in the chatroom and ask the system to regenerate the script accordingly. In Conditions B and C, users can also directly edit the step-by-step NL explanation.

\vspace{1mm}
\noindent{\bf \textit{Providing Feedback in the Chatroom.}} As shown in Figure~\ref{fig:likert_ratings}, 50\% of participants in Condition A, 66\% in Condition B, and 68\% in Condition C agreed or strongly agreed that providing feedback in the chatroom is effective for error correction. However, some participants complained that chatroom is not an efficient way to fix errors, since \new{it requires more clarifications to specify which step (P9), and type out every detail to fix even small errors (P15).} 


\vspace{1mm}
\noindent{\bf \textit{Editing Step-by-step NL Explanations.}} In Conditions B and C, {\tool} allowed users to fix errors by directly editing the step-by-step NL description of the scripts. As shown in Figure~\ref{fig:likert_ratings}, 61\% in Condition B and 66\% in Condition C agreed or strongly agreed that directly editing the step-by-step NL description is effective in fixing errors. Participants found that step-by-step NL explanations were valuable in multiple ways: \new{including clearly indicating where modifications were needed (P9), offering flexibility in program modification while illustrating the problem-solving sequence (P20), and helping verify whether their instructions were correctly interpreted (P16).} However, compared to directly providing feedback in the chatroom, we found no statistically significant differences in the mean ratings between these two features.

\subsubsection{ {\bf \textit{Participants' Perceptions of Different Program Explanation Features}}}
We experimented with two methods to explain a synthesized script. As described in Section~\ref{sec:conditions}, in Condition A, {\tool} simply generates an NL summary by prompting an LLM. In Conditions B and C, {\tool} uses a rule-based method to translate the synthesized script into a step-by-step NL explanation.

\vspace{1mm}
\noindent{\bf \textit{NL Summary.}} As shown in Figure~\ref{fig:likert_ratings}, in Condition A, only 47\% of participants agreed or strongly agreed that NL summary helps them understand the script and 53\% agreed or strongly agreed that NL summary accurately represented program behavior. Some participants found the NL summaries challenging to work with, noting that \new{they were overly verbose (P10), and required effort to parse and verify against original prompts (P18).} 

\vspace{1mm}
\noindent{\bf \textit{Step-by-Step NL Explanation.}} As shown in Figure~\ref{fig:likert_ratings}, significantly more users appreciated the step-by-step NL explanations compared with NL summary. 82\% in Condition B and 95\% in Condition C agreed or strongly agreed that these explanations helped them understand the scripts. Furthermore, 87\% in Condition B and 92\% in Condition C agreed or strongly agreed that the step-by-step NL explanations accurately represented program behavior.
Participants found that \new{the step-by-step approach enhanced program comprehension by improving readability for error checking (P4) and making scripts easier to understand and execute (P21).} 
The results suggest that structured step-by-step natural language explanations can effectively help users better validate and understand the scripts' behavior.

\begin{figure*}[t!]
    \centering
    \includegraphics[width=\textwidth]{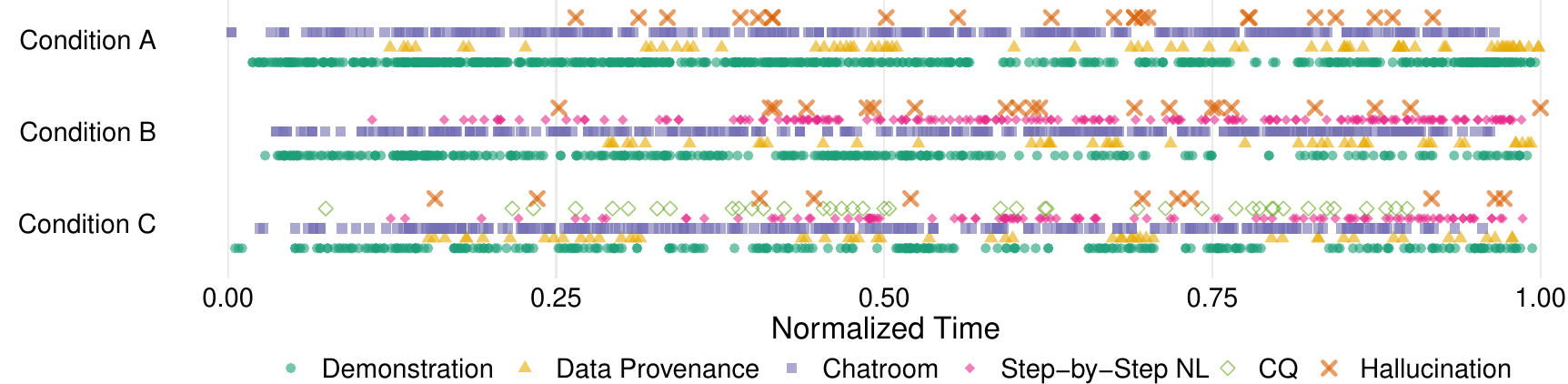}
    \captionsetup{font=small, labelfont=bf}
    \caption{A scatter plot displays events occurring throughout normalized time across all conditions.} 
    \label{fig:logged_events}
\end{figure*}

\subsubsection{ {\bf \textit{Participants' Perceptions of Data Provenance Feature.}}}

The majority of the participants appreciated the data provenance feature that allows them to check the contents of different versions. 66\% of participants in Condition A, 71\% in Condition B, and 68\% in Condition C agreed or strongly agreed that seeing data provenance helps check the contents of different versions of tables. Participants found that \new{the data provenance view helped build confidence by showing correspondences between selected tables (P15), and became more valuable as tasks increased in complexity (P16).} 

\subsubsection{\bf {\textit{Utility Rates of Different Features.}}}
\label{sec:utility}
To understand user behavior when using {\tool}, we analyzed our system logs and recordings and examined how participants interacted with different features to complete their tasks effectively. Figure~\ref{fig:logged_events} renders the temporal distribution of different user events for all task sessions in different conditions. Given the variances in the task completion time of each session, we normalized the occurrence of each user event with respect to the task session length. Notably, participants in Condition C relied less on user-initiated features like demonstrations and step-by-step explanations compared to Conditions A and B. The differences were statistically significant ($p$-values = 1.2e-02, and 5.64e-04, respectively). This indicates that participants in Condition C did not need to spend time on re-prompting and demonstrations. Instead, participants engaged more with multiple-choice clarification questions, which helped them effectively clarify their intent. This improvement aligns with the reduced completion times and improved success rate reported in Section~\ref{sec:performance}.

\subsection{RQ3: Reasons of {\tool}'s Effectiveness}

We analyzed participants' responses in the post-study survey and identified several key findings that explain why {\tool} helped participants better perform data wrangling tasks.

\vspace{1mm}
\noindent{\bf \textit{Finding 1: Step-by-step NL explanation helps users understand the program behavior.}} We found that the step-by-step NL explanation of the program accelerated the process of program comprehension.  82\% of participants in Condition B, and 95\% of participants in Condition C agreed or strongly agreed that step-by-step NL descriptions helped them understand the synthesis script, as shown in Figure~\ref{fig:likert_ratings}. P19 said, ``\textit{It divided the task into NL steps, easy to understand the steps performed.}'' 
In contrast, participants in Condition A spent more time comprehending the NL summary of the program. 
P8 said, ``\textit{I disliked that the natural language summaries didn't clearly distinguish between partially inaccurate and accurate responses. Their similarity made it time-consuming to read the entire summary and compare different versions.}''

\vspace{1mm}
\noindent{\bf \textit{Finding 2: Direct edits on step-by-step NL statement make correction easier.}} We observed that the LLM misinterpreted user intent in many cases and generated an incorrect script initially. In Condition A, participants could only provide feedback to fix incorrect scripts by providing more demonstrations or giving NL feedback in the chatroom. These two kinds of feedback turned out to be not very effective in guiding the LLM to refine the script.  Specifically, in Condition A, 57\% of participants encountered an incorrect script in their first attempt and spent an average of 4 minutes and 6 seconds to fix the incorrect script. By contrast, in Conditions B and C, participants could directly edit the NL description of an erroneous step of the script. This provided a more precise and fine-grained feedback to the LLM. While 49\% of participants in Condition B and 34\% of participants in Condition C encountered an incorrect script in their first attempts, they only spent 2 minutes and 41 seconds and and 2 minutes and 25 seconds to fix the incorrect script. Participants appreciated that they could easily edit and delete steps in the script (P7) and make corrections while preserving the conversational context (P1).

\vspace{1mm}
\noindent{\bf \textit{Finding 3: Clarification questions help users clarify their intent effectively.}} Participants frequently provided incomplete or ambiguous prompts. In Condition C, participants could directly answer relevant multiple-choice clarification questions (CQs) without needing to articulate their intent in a new prompt. A total of 40 CQs were generated in Condition C. Notably, 62\% of participants who answered CQs succeeded on their first attempt. By contrast, in Conditions A and B, only 43\% and 51\% of participants succeeded in their first attempt. In the post-task survey, 89\% of participants agreed or strongly agreed that answering CQs was an effective way to express their intent, as shown in Figure~\ref{fig:likert_ratings}. P14 commented, ``\textit{I appreciate how {\tool} asks clarifying questions before generating the program. This approach is better than directly outputting information, as ChatGPT might do, without truly understanding the user's intent.}'' Similarly, P11 stated, ``\textit{I appreciated the questions posed by the LLM, as they eliminated the need for me to specify all the details in my initial prompt.}''

\subsection{RQ4: Impact of User Expertise}
\subsubsection{ {\bf \textit{Overall Performance Comparison of Users with Different Expertise.}}}
We categorized participants into three groups based on their programming experiences, including novices (<1 year, N=12), intermediate programmers (1-5 years, N=13), and experts (> 5 years, N=12). We compared task completion time across these three groups. Experts completed tasks in an average of 4 minutes and 17 seconds, intermediate programmers in 5 minutes and 14 seconds, and novices in 5 minutes and 32 seconds. These mean differences were not statistically significant (ANOVA test: $p$-value=1.33e-01). The average number of attempts was similar across groups---experts made 1.67 attempts, intermediate programmers 1.74 attempts, and novices 1.67 attempts. These mean differences were also not statistically significant differences (ANOVA test: $p$-value=0.92e-01). This suggests that user expertise had a limited impact on performance. We interpret this as a positive impact of {\tool}, which narrows the performance gap between users of different levels of programming expertise .

\subsubsection{ {\bf \textit{Task-specific Comparison of Users with Different Expertise.}}}

In a detailed analysis of individual tasks, we found that task completion time was generally consistent across user groups for Tasks 1-6, aligning with our overall finding of no significant differences. However, Task 7 emerged as a notable exception. Experts completed this task in an average of 2 minutes and 10 seconds, while intermediate programmers and novices took considerably longer time---6 minutes and 54 seconds and 6 minutes and 53 seconds, respectively. This difference was statistically significant (ANOVA test p-value=1.15e-03). One plausible reason is that Task 7 required complex operations such as conditional data transformations and statistical testing. In particular, participants were required to perform a statistical test on their data and then apply an conditional filtering of data records based on the test results. Compared with novices and intermediate programmers, experts have a significant advantage in both knowledge and experience to handle such a complex task. This suggests that while {\tool} successfully leveled the playing field for common data wrangling operations, more complex tasks may still benefit from user expertise.

\begin{figure*}[t!]

    \centering
    \includegraphics[width=\textwidth]{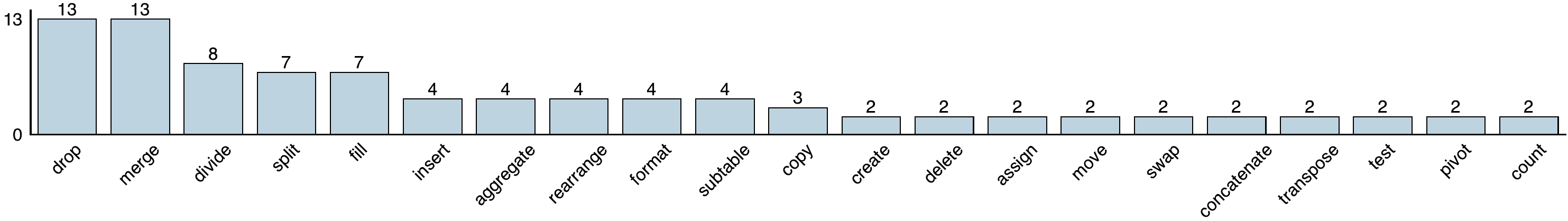}
    \captionsetup{font=small, labelfont=bf}
    \caption{The number of DSL statements that cover
different kinds of data-wrangling tasks in the case study.}  
    
    \label{fig:translation_rules}

\end{figure*}

\subsection{RQ5: Generalizability of {\tool}}

To evaluate the generalizability of {\tool}, we conducted a case study leveraging 10 data wrangling tasks collected from prior work~\cite{kandel2011wrangler, kasica2020table, li2024sheetcopilot, chen2022rigel}. Since the original tasks primarily focused on single-table tasks, we extended the benchmark by including another 14 multi-table data wrangling tasks introduced by Kasica et al.~\cite{kasica2020table}. The final benchmark includes a total of 24 tasks.\footnote{We provide the task descriptions and synthesized data wrangling scripts in the supplemental material.} It includes some common data wrangling tasks, such as transposing tables (benchmark ID 24), splitting one column into two columns based on a delimiter (benchmark ID 16), and summarizing tables by calculating average numbers (benchmark ID 21). It also includes more difficult tasks that involve conditional operations. For instance, deleting rows as long as there is a missing value in the row (benchmark ID 23). 
Since this quantitative study aims to verify that {\tool} can solve various data wrangling tasks instead of evaluating its learnability or usability, the first two authors independently solved these 24 tasks using {\tool}. They represent expert users who are familiar with the tool and have rich experience in data wrangling. Thus, the results of this study should be interpreted as the performance of {\tool} in ideal situations. A task is considered failed if an author cannot solve the task after 10 minutes. Overall, the two authors solved all 24 tasks with an average of 62.67 seconds and 75.90 seconds, respectively. On average, they required 1.54 and 1.33 attempts per task. This result shows that {\tool} can be used to solve a wide range of data wrangling tasks.

\subsubsection{ {\bf \textit{DSL Statement Distribution in Synthesized Programs.}}}
To understand the generalizability of individual DSL statements that are supported by {\tool}, we investigated the number of DSL statements that appeared in the synthesized programs.  Specifically, Figure~\ref{fig:translation_rules} shows how often each DSL statement was synthesized in these 24 tasks. The translation rules for \textsc{drop} and \textsc{merge} were the most utilized, both were triggered 13 times. The frequent use of \textsc{drop} implies that many data wrangling tasks require removing undesired data, e.g., rows with missing values, columns with sensitive information, etc. Furthermore, the frequent use of \textsc{merge} implies a common need of combining cleaned data from multiple tables into one single final table.

\subsubsection{ {\bf \textit{Effectiveness of Clarification Questions}}}
When solving these 24 tasks, the first author encountered 46 CQs, while the second author encountered 32 CQs. The authors examined the helpfulness of each CQ by checking if (1) it asked relevant and essential information critical to task success, and (2) it helped clarify their intent. The first author found 87\% of the questions helpful, while the second author rated 94\% as helpful. Regarding user responses to CQs, the first author selected the provided options 74\% of the time, while the second author selected the provided options 75\% of the time. In the remaining case, they chose ``Other (please specify)'' to clarify their intent. This implies  that the auto-generated options in the CQs are likely to cover user intent in the majority of cases.



\section{DISCUSSION}
\subsection{Design Implications}

\subsubsection{{\bf \textit{Aligning the LLM with  human intent via interactive clarification questions}}}
Compared to existing data wrangling tools, our tool allows the LLM to proactively ask clarification questions to engage with users and understand their intent. The success of {\tool} also echoes Grice's Maxims of Manner in cooperative principles~\cite{bernsen1996cooperativity}, which emphasize being clear and avoiding ambiguity in communication with users. Besides, we implement multiple-choice questions with potential options for users to select from, rather than relying on open-ended questions. This allows for precise intent clarification while reducing the manual effort of typing down details to articulate user intent. By integrating the Q\&A history, our tool improves the accuracy of data wrangling tasks and provides a more cooperative and user-friendly experience.

\subsubsection{{\bf \textit{Scaffolding LLM understanding using step-by-step NL explanations}}} The synthesized script can be interpreted as a manifesto of the LLM's understanding of user intent. If the LLM misinterprets user intent, the synthesized script is likely to be wrong. The NL explanation in {\tool} serves as an effective bi-directional communication vehicle between the user and the LLM. By explaining the script in NL, users can easily understand how the LLM interprets their intent without the need for programming knowledge. Additionally, the step-by-step structure of the NL explanation provides an effective scaffold for users to indicate the generation errors and provide more targeted and granular feedback than supplementing more demonstrations or entering feedback in the chatbot. The targeted and granular feedback can further help the LLM rectify its misunderstanding and refine the synthesized script.

\subsubsection{{\bf \textit{Help user track data lineage of multiple tables using data provenance}}}
{\tool} provides the data provenance feature that enables users to inspect different versions of their data. This is particularly useful for complex data wrangling tasks involving multiple tables. This feature allows users to trace the lineage of tables, revealing how specific data points have been modified throughout the wrangling process. Users can interact with the data provenance view by clicking on a table node, triggering the corresponding table content to appear in the table view panel. This visual correspondence, also known as \textit{brushing}~\cite{becker1987brushing}, allows users to visualize table interdependencies and build confidence in the synthesized script.

\subsection{Limitations}
First, our predefined DSL functions may not cover all possible data wrangling scenarios. The DSL does not explicitly include traditional control structures like \texttt{for} and \texttt{while} loops. Second, to mitigate hallucinations, we only implement a syntax checker to validate the synthesized script, omitting semantic correctness. 
Third, some participants mentioned the lack of advanced table features, such as the ability to use formulas directly in table cells, which are commonly supported by Microsoft Excel and Google Sheets. However, {\tool} currently does not support synthesizing formulas. Finally, our user study is conducted in a controlled lab setting. It does not reflect the complexity of real-world data wrangling scenarios, which may involve more complex operations, richer datasets, and even collaboration between multiple users in a longer period of time.  


\subsection{Future Work}

\subsubsection{{\bf \textit{Challenges and considerations of prompt design}}}
LLMs introduce several design challenges. First, LLMs are prone to hallucination~\cite{leiser2024hill}. We acknowledge that some errors or failures in our study stem from hallucinations. We observe that LLMs misuse our DSL even when our prompt clearly states the DSL grammar and provides detailed usage instructions. We have attempted to mitigate this by building a syntax checker to give feedback to the LLM. However, this increases synthesis time, and hallucination still exists. Second, we invested considerable effort and time to balance the level of detail and the number of few-shot examples in the prompt. We found too few examples might increase the error rate and elicit overgeneralized answers. Conversely, too many specific examples might reduce an LLM's inference ability. For instance, we observed that LLMs tend to rely more on pattern matching of the provided examples rather than truly understanding the underlying tasks. Future work should consider these prompt engineering challenges that have recently been coined as the \textbf{gulf of envision}~\cite{subramonyam2024bridging}.
 

\subsubsection{{\bf \textit{Handling ambiguous user intent}}}

We found many errors stem from users' ambiguous NL inputs and demonstrations. These errors cannot be solely attributed to the model's hallucination problems or prompt designs. When given ambiguous input, the LLM could interpret it differently. For example, if a user requests to ``sort the elements in the table,'' several questions arise: Should the sorting be ascending or descending? Should LLM sort it based on the name alphabetically or the ID values? Our study shows that simple clarification questions could greatly mitigate the effect of ambiguous intent. However, current LLM systems~\cite{achiam2023gpt, floridi2020gpt, anil2023palm, touvron2023llama} rarely check with users or help them clarify intent, instead synthesizing output in one shot. Future work could develop more interactive approaches to disambiguate users' intent, rather than expecting the LLM to magically ``read minds'' when given ambiguous input.

\subsubsection{{\bf \textit{Mitigating the abstract matching problem in NL interfaces}}}

We analyzed the recordings to understand why some users failed or spent more time to complete tasks. We found that while these users typically had well-formed intent, they often struggled to articulate them with sufficient precision and detail for the LLM to accurately map their intent to desired solutions. Such a challenge is common in NL interfaces, and it can be generally seen as the \textbf{gulf of execution}~\cite{hutchins1985direct}, or more precisely the \textbf{abstraction matching problem}~\cite{liu2023wants}: \textit{selecting words that align with the appropriate level of abstraction required by the system to choose correct actions.} However, our clarification question is not currently designed to bridge the abstraction gap. A potential solution is to build a bidirectional channel that maps the NL interface to the code. This could create visual correspondence, allowing users to link specific NL to code. Such a feature could also serve as a valuable learning tool for end-users.

\subsubsection{{\bf \textit{Visualizing intermediate states for fine-grained examination}}}
Some users have expressed concerns about the visibility of intermediate states in a data wrangling script. They noted that some intermediate states are not visible when executing the script. For example, when synthesizing a script to delete a column based on the p-value computed by a statistical test, users may want to check the p-value to make sure the t-test is performed correctly. There have been some existing efforts to enable users to inspect the intermediate states of a synthesized script~\cite{zhou2022intent, chen2023miwa, tian2024sqlucid}. A design challenge in the domain of data wrangling is that some intermediate states may involve a large volume of data, which may be overwhelming to visualize to users. Furthermore, given the large volume of data, helping users recognize the intermediate state change before and after a data transformation operation could be challenging. These challenges are worthwhile to investigate in the future.

\subsubsection{{\bf \textit{Supporting multimodal interaction paradigms}}}
{\tool} currently supports demonstrations and NL interaction. We believe recent advancements in multimodal LLMs \cite{hu2024bliva, li2023efficient, li2022blip} will further expand the potential design space for data wrangling tasks. For instance, future systems could enable users to sketch and circle data to clarify references such as ``that column'' or ``this row.'' They could also support hand gestures to zoom, filter, and navigate multi-table tasks that are tedious with NL prompts alone. However, implementing such systems poses several challenges. First, it is difficult to ``translate'' different intent from various modalities into a unified language. Second, a multimodal system should also be able to handle ambiguous inputs and generate conclusions with varying levels of certainty~\cite{jaimes2007multimodal}. 
Future work should consider these challenges, also known as  \textit{fusion}~\cite{jaimes2007multimodal} issues, in the era of multimodal LLMs.



\section{CONCLUSION}

This paper introduces {\tool}, a mixed-initiative LLM system for data wrangling. Compared with previous work, {\tool} allows the LLM to proactively ask clarification questions to resolve ambiguities in user intent. It also provides multiple feedback mechanisms to help the user understand and refine the synthesized script. A within-subjects study (n = 38) demonstrates that {\tool}'s features significantly improved data wrangling efficiency. Furthermore, a case study with 24 additional tasks demonstrates the generalizability of {\tool}'s effectiveness on different kinds of tasks.


\bibliographystyle{ACM-Reference-Format}




\onecolumn
\appendix


\section{Domain Specific Language}
\label{appendix:dsl}


\begin{table*}[h]
\small
\begin{tabular}{p{37pc}}
\hline
     \textbf{\cellcolor{gray!17}{Table-level Functions}}\\
\hline
   1. create\_table(row\_number, column\_number): Returns an empty table with the specified number of rows and columns.\\
   2. delete\_table(table\_name): Deletes a table from the database.\\
   3. pivot\_table(table, index, columns, values, aggfunc): Reshapes the table so that each unique value in columns becomes a separate column, with index values as row headers, and the corresponding values filled in their respective cells.\\
   4. merge(table\_a, table\_b, how='outer', on=None): Merges two tables based on a common column.\\
   5. subtable(table, labels, axis): Extracts a subtable from a DataFrame based on specified rows or columns.\\
   6. transpose(table): Transposes the given table.\\
   \textbf{\cellcolor{gray!17}{Column/Row-level Functions}}\\
   1. insert(table, index, index\_name, axis): Inserts an empty row or column at the specified index in the table. Other rows or columns will be moved down or to the right.\\
   2. drop(table, label, axis): Drops one or more rows or columns from the table.\\
   3. assign(table, start\_row\_index, end\_row\_index, start\_column\_index, end\_column\_index, values): Assigns fixed constant values to specific cells in the table.\\
   4. move(origin\_table, origin\_index, target\_table, target\_index, axis): Moves a row or column from the origin table to the target table.\\
   5. copy(origin\_table, origin\_label, target\_table, target\_label, axis): Copies a row or column from the origin table to the target table at the specified label.\\
   6. swap(table\_a, label\_a, table\_b, label\_b, axis): Swaps rows or columns between two tables.\\
   7. rearrange(table, by\_values=None, by\_array=None, axis): Rearranges the rows or columns of the table based on the specified order.\\
   8. divide(table, by, axis): Divides the table by the specified row or column, returning a list of tables.\\
   9. fill(table, method, labels, axis): Fills missing values in the table using the specified method.\\
   \textbf{\cellcolor{gray!17}{Summarization Functions}}\\
   1. aggregate(table, functions, axis): Aggregates the table using a specified function.\\
   2. test(table\_a, label\_a, table\_b, label\_b, strategy, axis): Compares two labels using the specified statistical test and returns a tuple (statistic, p\_value).\\
   3. count(table, label, value, axis): Counts the occurrences of a specified value within a given column or row in a DataFrame, then stores the result in a new DataFrame.\\
   \textbf{\cellcolor{gray!17}{String Operation Functions}}\\
   1. concatenate(table, label\_a, label\_b, glue, new\_label, axis): Concatenates two rows or columns using a string as glue and appends the merged row or column to the table.\\
   2. split(table, label, delimiter, new\_label\_list, axis): Separates rows or columns based on a string delimiter within the values.\\
   3. format(table, label, pattern, replace\_with, axis): Formats the values in a row or column based on the specified pattern and "replace\_with" using \texttt{re.sub()}.\\
\bottomrule
\label{tab:DSLtable}
\caption{
   DSL of {\tool}.
} \\
\end{tabular}
\end{table*}
\vspace{2em}  
\clearpage


\section{Prompt Designs}
\label{appendix:prompt}


\begin{table*}[h!]
\small
\begin{tabular}{p{0.98\textwidth}}
\toprule
   \textbf{\cellcolor{gray!17}{CONTEXT}}\\
   You are a professional data scientist. A user has made some changes in the CSV files.\\
   Your task is to understand the user intent regarding how they want to clean the data.\\
   \textbf{\cellcolor{gray!17}{OBJECTIVE}}\\
   Ask clarification questions to understand the user intent.\\
\textbf{\cellcolor{gray!17}{GUIDELINES}}\\
   1. Infer their intent through the table diff and user instruction. Do not infer beyond the information provided in the input.\\
   2. Avoid directly asking, "What is your intent?" Instead, ask questions related to the changes made in the table and the instructions given.\\
   3. If the user intent is clear, you can output the intent summary.\\
\textbf{\cellcolor{gray!17}{INPUT}}\\
   - Sheet Information: the name of the sheet, the headers, and the number of rows in the table.\\
   - Table Diff: the changes made to the table.\\
   - User Instruction: the user's instruction that indicates the changes they want to make.\\
   \textbf{\cellcolor{gray!17}{OUTPUT}}\\
   Your output must be in JSON list form.\\
   If you need more information, output a question to ask the user:\\
   \{\\
       \qquad"type": "question",\\
       \qquad"summary": "<summary>",\\
       \qquad"question": "<question>",\\
       \qquad"choices": ["<choice\_1>", "<choice\_2>", ..., "<choice\_n>", "other"]\\
   \}\\
   If the intent is clear enough, output a summary of the user intent:\\
   \{\\
      \qquad"type": "finish",\\
      \qquad"summary": "<summary>"\\
   \}\\
  \textbf{\cellcolor{gray!17}{EXAMPLES}}\\
   \colorbox[HTML]{ffe5cc}{<SOME EXAMPLES>}\\
\bottomrule
\caption{
   The prompt of intent summarization \& CQ generation.
}
\label{tab:intent_summarization_1}
\end{tabular}
\end{table*}
\clearpage



\begin{table*}[h!]
\small
\begin{tabular}{p{0.98\textwidth}}
\toprule
\textbf{\cellcolor{gray!17}{CONTEXT}}\\
You are a professional data scientist. You have already asked some clarification questions and the user has replied.\\
Now, you might want to ask additional questions to gain a deeper understanding of their intent.\\
\textbf{\cellcolor{gray!17}{OBJECTIVE}}\\
\colorbox[HTML]{ffe5cc}{<SAME AS TABLE \ref{tab:intent_summarization_1}>}\\
\textbf{\cellcolor{gray!17}{GUIDELINES}}\\
\colorbox[HTML]{ffe5cc}{<SAME AS TABLE \ref{tab:intent_summarization_1}>}\\
 
\textbf{\cellcolor{gray!17}{INPUT}}\\
- Sheet Information:  the name of the sheet, the headers, and the number of rows in the table.\\
- Table Diff: the changes made to the table, including the cells that have been modified.\\
- User Instruction: the user's instruction that indicates the changes they want to make.\\
- Chat History: The history of chat between the assistant and the user.\\
\textbf{\cellcolor{gray!17}{OUTPUT}}\\
\colorbox[HTML]{ffe5cc}{<SAME AS TABLE \ref{tab:intent_summarization_1}>}\\
\textbf{\cellcolor{gray!17}{EXAMPLES}}\\
\colorbox[HTML]{ffe5cc}{<SOME EXAMPLES>}\\
\bottomrule
\caption{
    The follow-up prompt of intent summarization \& CQ generation.
    \label{tab:intent_summarization_2}
}
\end{tabular}
\end{table*}



\begin{table*}[h!]
\small
\begin{tabular}{p{0.98\textwidth}}
\toprule
\textbf{\cellcolor{gray!17}{CONTEXT}}\\
You are a professional data scientist.\\
Your task is to generate a step-by-step plan to clean the data based on the user intent and the sheet information.\\
\textbf{\cellcolor{gray!17}{OBJECTIVE}}\\
Generate a step-by-step plan to clean the data based on the user intent and the sheet information.\\
\textbf{\cellcolor{gray!17}{DSL GRAMMAR}}\\
\colorbox[HTML]{ffe5cc}{<DSL\_GRAMMAR>}\\
\textbf{\cellcolor{gray!17}{GUIDELINES}}\\
1. You should specify the DSL function after the description. Do not add or invent new functions.\\
2. You should point out the arguments for each function based on the given description, please refer to the column and row information in the sheet information.\\
3. Please attention that the row index starts from 0, which is the header row. The column index starts from 1.\\
\textbf{\cellcolor{gray!17}{INPUT}}\\
- Sheet Information: the name of the sheet, the headers, and the number of rows in the table.\\
- User Intent: the user intent.\\
\textbf{\cellcolor{gray!17}{OUTPUT}}\\
- Step-by-step plan: a JSON list.\\
    \qquad- Each step should only include the function name and its description.\\
    \qquad- If there are multiple steps, list them in the order they should be executed.\\
\textbf{\cellcolor{gray!17}{EXAMPLES}}\\
\colorbox[HTML]{ffe5cc}{<SOME EXAMPLES>}\\
\bottomrule
\caption{
    \label{prompt:plan}
    The prompt for step-by-step planning.
}
\end{tabular}
\end{table*}



\begin{table*}[h]
\small
\begin{tabular}{p{0.98\textwidth}}
\toprule
\textbf{\cellcolor{gray!17}{CONTEXT}}\\
You are a professional data scientist.\\
Your task is to generate a step-by-step plan to clean the data based on the user intent and an error message from the last step-by-step plan.\\
\textbf{\cellcolor{gray!17}{OBJECTIVE}}\\
\colorbox[HTML]{ffe5cc}{<SAME AS TABLE \ref{prompt:plan}>}\\
\textbf{\cellcolor{gray!17}{DSL GRAMMAR}}\\
\colorbox[HTML]{ffe5cc}{<DSL\_GRAMMAR>}\\
\textbf{\cellcolor{gray!17}{GUIDELINES}}\\
\colorbox[HTML]{ffe5cc}{<SAME AS TABLE \ref{prompt:plan}>}\\
\textbf{\cellcolor{gray!17}{INPUT}}\\
- Sheet Information: the name of the sheet, the headers, and the number of rows in the table.\\
- User Intent: the user intent.\\
- Last step-by-step plan: A JSON list. Each step should include the function name and its description.\\
- Error Message: the error message from the last generation.\\
\textbf{\cellcolor{gray!17}{OUTPUT}}\\
\colorbox[HTML]{ffe5cc}{<SAME AS TABLE \ref{prompt:plan}>}\\
\textbf{\cellcolor{gray!17}{EXAMPLES}}\\
\colorbox[HTML]{ffe5cc}{<SOME EXAMPLES>}\\
\bottomrule
\caption{
    \label{prompt:plan_with_error}
    The prompt for step-by-step planning with error messages.
}
\end{tabular}
\end{table*}



\begin{table*}[h]
\small
\begin{tabular}{p{0.98\textwidth}}
\toprule
\textbf{\cellcolor{gray!17}{CONTEXT}}\\
You are a professional DSL (Domain Specific Language) generator.\\
You will be given a step-by-step description of a data cleaning plan.\\
You need to follow the description and create a DSL script to help user clean and manipulate the data.\\
\textbf{\cellcolor{gray!17}{OBJECTIVE}}\\
Create a DSL script to clean the data based on the description.\\
\textbf{\cellcolor{gray!17}{DSL GRAMMAR}}\\
\colorbox[HTML]{ffe5cc}{<DSL\_GRAMMAR>}\\
\textbf{\cellcolor{gray!17}{GUIDELINES}}\\
1. You should only use the DSL functions provided in the DSL Grammar. Do not add or invent new functions.\\
2. For every step, you need to find the best function from the DSL Grammar to perform the described action.\\
3. Table names should end with ".csv" to indicate that they are CSV files.\\
4. If you use an integer for a row index, it should be 0-based. If you use a string for a row index, it should start from "1".\\
5. For None values, you can use "null" in the output.\\
6. This DSL script is not the final program. Please use the table names instead of real pandas DataFrames in the arguments.\\
\textbf{\cellcolor{gray!17}{INPUT}}\\
- Sheet information: the name of the sheet, the headers, and the number of rows in the table.\\
- Step-by-step Plan: a detailed description of the process.\\
\textbf{\cellcolor{gray!17}{OUTPUT}}\\
Your output should be in a JSON form.\\
The JSON should contain two parts:\\
- "required\_tables": A list of table names that are required to perform the cleaning process.\\
- "program": A list of objects that represent the functions to be applied to the tables. Each object of the list should contain the function name and its arguments. If a function needs to be applied to special cells, you can add a "condition" parameter to the object.\\
Do not add any other characters to the output.\\
\textbf{\cellcolor{gray!17}{EXAMPLES}}\\
\colorbox[HTML]{ffe5cc}{<SOME EXAMPLES>}\\
\bottomrule
\caption{
    \label{prompt:generate}
    The prompt for generating a DSL script.
}
\end{tabular}
\end{table*}



\begin{table*}[h]
\small
\begin{tabular}{p{0.98\textwidth}}
\toprule
\textbf{\cellcolor{gray!17}{CONTEXT}}\\
You are a professional DSL (Domain Specific Language) generator.\\
You will be given a step-by-step description of a data cleaning plan, and an error message from the last generation.\\
You need to follow the description and create a DSL script to help the user clean and manipulate the data.\\
\textbf{\cellcolor{gray!17}{OBJECTIVE}}\\
\colorbox[HTML]{ffe5cc}{<SAME AS TABLE \ref{prompt:generate}>}\\
\textbf{\cellcolor{gray!17}{DSL GRAMMAR}}\\
\colorbox[HTML]{ffe5cc}{<DSL\_GRAMMAR>}\\
\textbf{\cellcolor{gray!17}{GUIDELINES}}\\
\colorbox[HTML]{ffe5cc}{<SAME AS TABLE \ref{prompt:generate}>}\\
\textbf{\cellcolor{gray!17}{INPUT}}\\
- Sheet information: the name of the sheet, the headers, and the number of rows in the table.\\
- Step-by-step Plan: a detailed description of the process.\\
- Error Message: The error message from the last generation.\\
\textbf{\cellcolor{gray!17}{OUTPUT}}\\
\colorbox[HTML]{ffe5cc}{<SAME AS TABLE \ref{prompt:generate}>}\\
\textbf{\cellcolor{gray!17}{EXAMPLES}}\\
\colorbox[HTML]{ffe5cc}{<SOME EXAMPLES>}\\
\bottomrule
\caption{
    \label{prompt:generate_dsl_with_error_message}
    The prompt for generating a DSL script with error messages.
}
\end{tabular}
\end{table*}



\begin{table*}[h]
\small
\begin{tabular}{p{0.98\textwidth}}
\toprule
\textbf{\cellcolor{gray!17}{CONTEXT}}\\
You are a professional DSL (Domain Specific Language) expert.\\
You will be given a required tables list, a DSL functions list in a JSON list format, and user intent.\\
You need to create a Python code snippet that executes the given DSL function.\\
\textbf{\cellcolor{gray!17}{OBJECTIVE}}\\
Create a Python code snippet that executes the given DSL functions list.\\
\textbf{\cellcolor{gray!17}{DSL GRAMMAR}}\\
\colorbox[HTML]{ffe5cc}{<DSL\_GRAMMAR>}\\
\textbf{\cellcolor{gray!17}{GUIDELINES}}\\
1. All output should include the \texttt{save\_table} function.\\
2. When creating a new table using the \texttt{save\_table} function, ensure the table is formatted as "\{name\}\_v\{version\}.csv". Please start the version from 0.\\
3. When using \texttt{merge\_table} function, here is the naming convention for the tables:\\
- merge\_table function: "merged\_v0.csv"\\
4. The row indexes are strings that start from "1" and should be enclosed in double-quotes.\\
5. Notice the number of output arguments for each function and assign them accordingly.\\
\textbf{\cellcolor{gray!17}{INPUT}}\\
- Required Tables: A list of table names that are required to perform the DSL functions.\\
- DSL Program: A JSON list containing the DSL functions to be executed.\\
- User Intent: A natural language description of the user intent.\\
\textbf{\cellcolor{gray!17}{OUTPUT}}\\
Your output should be between \`{}\`{}\`{} tags and contain the Python code snippet that executes the given DSL function.\\
\textbf{\cellcolor{gray!17}{EXAMPLES}}\\
\colorbox[HTML]{ffe5cc}{<SOME EXAMPLES>}\\
\bottomrule
\caption{
    \label{prompt:execute}
    The prompt for executing a DSL script.
}
\end{tabular}
\end{table*}



\begin{table}[h]
\small
\begin{tabular}{p{0.98\textwidth}}
\toprule
\textbf{\cellcolor{gray!17}{CONTEXT}}\\
You are a professional DSL (Domain Specific Language) generator.\\
You will be given an instruction to create a DSL and information including a previous version DSL list and a DSL grammar.\\
\textbf{\cellcolor{gray!17}{OBJECTIVE}}\\
Create the DSL script as the instructions specified.\\
\textbf{\cellcolor{gray!17}{DSL GRAMMAR}}\\
\colorbox[HTML]{ffe5cc}{<DSL\_GRAMMAR>}\\
\textbf{\cellcolor{gray!17}{GUIDELINES}}\\
1. You should only use the DSL functions provided in the DSL Grammar. Do not add or invent new functions.\\
2. Table names should end with ".csv" to indicate that they are CSV files.\\
3. This DSL script is not the final program. Please use the table names instead of real pandas DataFrames in the arguments.\\
\textbf{\cellcolor{gray!17}{INPUT}}\\
- Previous generated DSL\\
- New Instruction\\
\textbf{\cellcolor{gray!17}{OUTPUT}}\\
Your output should be in a JSON object. Each object should contain the function name and its arguments.\\
If a function needs to be applied to specific cells, you can add a "condition" parameter to the object.\\
\textbf{\cellcolor{gray!17}{EXAMPLES}}\\
\colorbox[HTML]{ffe5cc}{<SOME EXAMPLES>}\\
\bottomrule
\caption{
    \label{prompt:update_dsl}
    The prompt for regenerating a DSL script based on user feedback.
}
\end{tabular}
\end{table}



\begin{table*}[h]
\small
\begin{tabular}{p{0.98\textwidth}}
\toprule
\textbf{\cellcolor{gray!17}{CONTEXT}}\\
You are a professional DSL (Domain Specific Language) generator.\\
You will be given instructions on how to change a DSL from the previous version to the new DSL.\\
\textbf{\cellcolor{gray!17}{OBJECTIVE}}\\
Change the DSL script as the instructions specified.\\
\textbf{\cellcolor{gray!17}{DSL GRAMMAR}}\\
\colorbox[HTML]{ffe5cc}{<DSL\_GRAMMAR>}\\
\textbf{\cellcolor{gray!17}{GUIDELINES}}\\
1. You should only use the DSL functions provided in the DSL Grammar. Do not add or invent new functions.\\
2. Table names should end with ".csv" to indicate that they are CSV files.\\
3. This DSL script is not the final program. Please use the table names instead of real pandas DataFrames in the arguments.\\
\textbf{\cellcolor{gray!17}{INPUT}}\\
- Previous generated DSL\\
- New Instruction\\
\textbf{\cellcolor{gray!17}{OUTPUT}}\\
Your output should be in a JSON object. Each object should contain the function name and its arguments.\\
If a function needs to be applied to specific cells, you can add a "condition" parameter to the object.\\
\textbf{\cellcolor{gray!17}{EXAMPLES}}\\
\colorbox[HTML]{ffe5cc}{<SOME EXAMPLES>}\\
\bottomrule
\caption{
    \label{prompt:edit_dsl}
    The prompt for editing a DSL script based on user feedback.
}
\end{tabular}
\end{table*}



\begin{table*}[h]
\small
\begin{tabular}{p{0.98\textwidth}}
\toprule
\textbf{\cellcolor{gray!17}{CONTEXT}}\\
You are a professional data scientist.\\
You will be given a DSL script that is used to clean and manipulate the data and a previous user intent.\\
\textbf{\cellcolor{gray!17}{OBJECTIVE}}\\
Update the user intent based on the new DSL script.\\
\textbf{\cellcolor{gray!17}{INPUT}}\\
- New DSL Script: a list of objects that represent the functions to be applied to the tables.\\
\textbf{\cellcolor{gray!17}{OUTPUT}}\\
New user intent based on the new DSL script.\\
\textbf{\cellcolor{gray!17}{EXAMPLES}}\\
\colorbox[HTML]{ffe5cc}{<SOME EXAMPLES>}\\
\bottomrule
\caption{
    \label{prompt:update_intent}
    The prompt for updating the user intent for program refinement.
}
\end{tabular}
\end{table*}



\begin{table*}[h]
\small
\begin{tabular}{p{0.98\textwidth}}
\toprule
\textbf{\cellcolor{gray!17}{CONTEXT}}\\
You are a professional Data Scientist.\\
You will be given a DSL script that is used to clean and manipulate the data.\\
\textbf{\cellcolor{gray!17}{OBJECTIVE}}\\
Provide a natural language description for the DSL script.\\
\textbf{\cellcolor{gray!17}{DSL GRAMMAR}}\\
\colorbox[HTML]{ffe5cc}{<DSL\_GRAMMAR>}\\
\textbf{\cellcolor{gray!17}{INPUT}}\\
- DSL Script: a list of objects that represent the functions to be applied to the tables.\\
\textbf{\cellcolor{gray!17}{OUTPUT}}\\
- NL Explanation: a natural language description of the DSL script.\\
\textbf{\cellcolor{gray!17}{EXAMPLES}}\\
\colorbox[HTML]{ffe5cc}{<SOME EXAMPLES>}\\
\bottomrule
\caption{
    \label{prompt:dsl_to_nl}
    The prompt to generate a NL summary of a synthesized DSL in Condition A of the user study.
}
\end{tabular}
\end{table*}


\clearpage
\section{Translation Rules}
\label{appendix:translation_rules}


\begin{table*}[!ht]
\Large
\renewcommand{\arraystretch}{2.5}  
\centering
\resizebox{\textwidth}{!}{
\begin{tabular}{|l|l|}
\hline
\textbf{Statements} & \textbf{Translation rules} \\
\hline
create\_table(X, Y) & ``Create a blank table with'' + X + ``rows and'' + Y + ``columns'' \\
\hline
delete\_table(X) & ``Delete the table'' + X \\
\hline
insert(table, X, Y, axis) & ``Insert a column at position'' + X + ``in the given table(s)'' \\
\hline
drop(table, X, axis) & ``Drop the column'' + X + ``in the given table(s)'' \\
\hline
assign(table, row, col, value, values) & ``Assign the values'' + json.dumps(values) + ``in the given table(s)'' \\
\hline
move(table, X, Y, axis) & ``Move the column'' + X + ``to column'' + Y + ``in the given table(s)'' \\
\hline
copy(table, X, Y, axis) & ``Copy the column'' + X + ``to column'' + Y + ``in the given table(s)'' \\
\hline
swap(table, X, Y, axis) & ``Swap the column'' + X + ``and the column'' + Y + ``in the given table(s)'' \\
\hline
merge(table\_a, X, Y) & ``Merge the given table(s) with the table'' + X + ``based on the values in the column'' + Y \\
\hline
concatenate(table, X, Y, Z, axis) & ``Concatenate the columns'' + X + ``and'' + Y + ``in the given table(s) with the glue'' + Z \\
\hline
split(table, X, Y, axis) & ``Split the values in the column'' + X + ``in the given table(s) with the delimiter'' + Y \\
\hline
transpose(table) & ``Transpose the given table(s)'' \\
\hline
aggregate(table, X) & ``Aggregate the given table(s) with the functions'' + X \\
\hline
test(A, B, C, D, E, axis) & ``Test the columns'' + B + ``in table'' + A + ``and'' + D + ``in table'' + C + ``using the'' + E \\
\hline
rearrange(table, X, axis) & ``Rearrange the columns in the given table(s) based on the values in the column'' + X \\
\hline
format(table, W, X, Y, axis) & ``Format the values in the column'' + W + ``in the given table(s) with the pattern'' + X + ``and replace them with'' + Y \\
\hline
divide(table, X, axis) & ``Divide the given table(s) by the values in the column'' + X \\
\hline
fill(table, X, Y) & ``Fill the missing values in the column'' + X + ``in the given table(s) with the method'' + Y \\
\hline
pivot\_table(table, W, X, Y, Z) & ``Create a pivot table in the given table(s) with the index'' + W + ``, columns'' + X + ``, values'' + Y + ``, and the aggregation function'' + Z \\
\hline
subtable(table, X, axis) & ``Extract a subtable from the given table(s) based on the columns'' + X \\
\hline
count(table, X, Y, axis) & ``Count the occurrences of the value'' + Y + ``in the column'' + X + ``in the given table(s)'' \\
\hline
\end{tabular}
}
\caption{Translation rules for DSL to Natural Language.}
\label{table:dsl_translation}
\end{table*}


\clearpage
\section{User Study Material}
\label{appendix:userstudy}
\subsection{Participant Demographics}


\begin{table*}[htbp]
\small
\setlength{\tabcolsep}{4pt}
\renewcommand{\arraystretch}{1.2}
\begin{tabular}{>{\centering\arraybackslash}m{0.8cm}>{\centering\arraybackslash}m{1.1cm}>{\centering\arraybackslash}m{2.2cm}>{\centering\arraybackslash}m{1.6cm}>{\centering\arraybackslash}m{3.8cm}>{\centering\arraybackslash}m{2.6cm}}
\hline
\textbf{ID} & \textbf{Gender} & \textbf{Programming Experience} & \textbf{Education} & \textbf{Department} & \textbf{Expertise} \\
\hline
P1 & Male& 1-5 years& Undergraduate& Computer Science& Novice\\
P2 & Male& 5+ years& PhD& Computer Science& Expert\\
P3 & Female& 1-5 years& PhD& Statistics& Novice\\
P4 & Male& 5+ years& Master& Computer Science& Expert\\
P5 & Male& 5+ years& Master& Computer Science& Expert\\
P6 & Female& 1-5 years& Undergraduate& Computer Science& Novice\\
P7 & Male& 1-5 years& Undergraduate& Computer Science& Novice\\
P8 & Male& 1-5 years& Undergraduate& Computer Science& Novice\\
P9 & Female& 1-5 years& PhD& Industrial Engineering& Novice\\
P10 & Female& < 1 year& PhD& Hospitality& End-user\\
P11 & Male& 5+ years& PhD& Computer Science& Expert\\
P12 & Male& 5+ years& PhD& Computer Science& Expert\\
P13 & Female& 1-5 years& PhD& Computer Science& Novice\\
P14 & Male& 1-5 years& Undergraduate& Computer Science& Novice\\
P15 & Male& 1-5 years& Undergraduate& Computer Science& Novice\\
P16 & Male& 1-5 years& Undergraduate& Computer Science& Novice\\
P17 & Male& 5+ years& PhD& Computer Science& Expert\\
P18 & Male& 5+ years& PhD& Computer Science& Expert\\
P19 & Female& 5+ years& Master& Computer Science& Expert\\
P20 & Male& 1-5 years& PhD& Physics& Novice\\
P21 & Male& 1-5 years& Undergraduate& Computer Science& Novice\\
P22 & Male& 5+ years& Undergraduate& Computer Science& Expert\\
P23 & Male& 5+ years& Undergraduate& Computer Science& Expert\\
P24 & Female& 5+ years& Undergraduate& Computer Science& Expert\\
P25 & Female& 5+ years& Undergraduate& Computer Science& Expert\\
P26 & Male& < 1 year& PhD& Materials Engineering& End-user\\
P27 & Male& < 1 year& Undergraduate& Game Design& End-user\\
P28 & Male& < 1 year& Master& Business Analysis& End-user\\
P29 & Female& 1-5 years& Undergraduate& Business Analysis& Novice\\
P30 & Male& < 1 year& PhD& Industrial Engineering& End-user\\
P31 & Male& < 1 year& Postdoc& Agricultural Engineering& End-user\\
P32 & Male& < 1 year& PhD& Game Design& End-user\\
P33 & Female& < 1 year& Undergraduate& Agricultural Economics& End-user\\
P34 & Male& 1-5 years& PhD& Mechanical Engineering& Novice\\
P35 & Male& < 1 year& PhD& Chemical Engineering& End-user\\
P36 & Male& < 1 year& PhD& Mechanical Engineering& End-user\\
P37 & Male& < 1 year& Undergraduate& Mechanical Engineering& End-user\\
P38 & Female& < 1 year& PhD& Communication& End-user\\
\hline
\end{tabular}
\caption{Participant Demographics}
\label{tab:demographics}
\end{table*}


\clearpage
\subsection{User Study Tasks}


\begin{table*}[htbp]
\centering
\renewcommand{\arraystretch}{1.2}
\begin{tabular}{l l p{0.6\textwidth}}
\toprule
\textbf{Task} & \textbf{Category} & \textbf{Description} \\
\midrule
Task 1 & Table Segmentation & Create four tables: ``Q1'' for sales from January to March, ``Q2'' for April to June, ``Q3'' for July to September, and ``Q4'' for October to December. \\
\addlinespace[0.5em]
Task 2$^{\ast}$ & Data Imputation & Fill all missing values in the rows across the table with the mean of each respective column. \\
\addlinespace[0.5em]
Task 3 & Categorical Analysis & Split the table into separate tables based on the ``Educational Level'' column. For each resulting table, summarize the ``income'' values by calculating their mean. \\
\addlinespace[0.5em]
Task 4 & Table Integration & Merge two tables based on matching student IDs, keeping only rows where the ID appears in both tables. Then count the number of ``Male'' entries in the ``sex'' column of the resulting table. \\
\addlinespace[0.5em]
Task 5 & Fuzzy Matching & Merge two tables using fuzzy matching on student names. Separate the merged names into distinct ``last name'' and ``first name'' columns. Finally, sort the resulting table alphabetically by last name. \\
\addlinespace[0.5em]
Task 6$^{\ast}$ & Data Cleaning & Remove rows with excessive missing values (user decides the threshold). For any remaining rows with missing values, fill these entries with the average values of their respective columns. \\
\addlinespace[0.5em]
Task 7 & Statistical Analysis & Conduct a statistical analysis comparing the ``Years of Experience'' column in Table 1 with the ``Age'' column in Table 2. If the result is statistically significant, remove one of these columns, retaining only the other. \\
\bottomrule
\addlinespace[0.3em]
\multicolumn{3}{l}{\footnotesize{$^{\ast}$Single-table tasks}} \\
\end{tabular}
\caption{User Study Tasks Overview}
\label{tab:data-tasks}
\end{table*}


\clearpage
\subsection{Post-task questionnaires}


\setlength{\belowcaptionskip}{10pt}  
\begin{table*}[htbp]
\centering
\small
\begin{tabular}{p{0.08\textwidth}p{0.72\textwidth}p{0.2\textwidth}}
\toprule
\textbf{ID} & \textbf{Questions} & \textbf{Scale} \\
\midrule
\multicolumn{3}{l}{{\bf User Confidence}} \\
Q1 & I felt confident about the final generated data cleaning script. & 1 (Strongly Disagree) -- 7 (Strongly Agree) \\[0.5em]
\midrule
\multicolumn{3}{l}{{\bf NASA TLX Questions}} \\
Q2 & How mentally demanding was using this tool? & 1 (Very Low) -- 7 (Very High) \\
Q3 & How hurried or rushed were you during the task? & 1 (Very Low) -- 7 (Very High) \\
Q4 & How successful would you rate yourself in accomplishing the task? & 1 (Failure) -- 7 (Perfect) \\
Q5 & How hard did you have to work to accomplish your level of performance? & 1 (Very Low) -- 7 (Very High) \\
Q6 & How insecure, discouraged, irritated, stressed, and annoyed were you? & 1 (Very Low) -- 7 (Very High) \\[0.5em]
\midrule
\multicolumn{3}{l}{{ \bf Ratings of Key features}} \\
Q7 & Demonstrating on the uploaded tables is a convenient way to express my intent. & 1 (Strongly Disagree) -- 7 (Strongly Agree) \\
Q8 & Using natural language in the chatroom is a convenient way to express my intent. & 1 (Strongly Disagree) -- 7 (Strongly Agree) \\
Q9\textsuperscript{§} & Answering clarification questions is an effective way to clarify my intent. & 1 (Strongly Disagree) -- 7 (Strongly Agree) \\
Q10 & Providing feedback through the chatroom is an effective way to fix errors. & 1 (Strongly Disagree) -- 7 (Strongly Agree) \\
Q11 & Directly editing descriptions of erroneous steps is an effective way to fix errors. & 1 (Strongly Disagree) -- 7 (Strongly Agree) \\
Q12\textsuperscript{‡} & Step-by-step NL descriptions helped me understand and validate script behavior. & 1 (Strongly Disagree) -- 7 (Strongly Agree) \\
Q13\textsuperscript{‡} & Step-by-step NL descriptions accurately represented the program's behavior. & 1 (Strongly Disagree) -- 7 (Strongly Agree) \\
Q14\textsuperscript{†} & NL summary of the script helped me understand and validate script behavior. & 1 (Strongly Disagree) -- 7 (Strongly Agree) \\
Q15\textsuperscript{†} & NL summary of the script accurately represented the program's behavior. & 1 (Strongly Disagree) -- 7 (Strongly Agree) \\
Q16 & Seeing table's data provenance and version contents was helpful. & 1 (Strongly Disagree) -- 7 (Strongly Agree) \\[0.5em]
\midrule
\multicolumn{3}{l}{{ \bf Open-ended Feedback}} \\
Q17 & What do you like about {\tool} in this setting? & Open-ended \\
Q18 & What do you dislike about {\tool} in this setting? & Open-ended \\
Q19 & What suggestions do you have for improving {\tool} in this setting? & Open-ended \\
\bottomrule
\\
\multicolumn{3}{l}{\small †: Condition A only; ‡: Condition B, C only; §: Condition C only} \\
\end{tabular}
\caption{Post-task survey questions including NASA TLX measures for subjective workload assessment}
\label{tab:posttask}
\end{table*}


\subsection{Post-study questionnaires}


\setlength{\belowcaptionskip}{10pt}  
\begin{table*}[htbp]
\centering
\small
\begin{tabular}{p{0.08\textwidth}p{0.72\textwidth}p{0.2\textwidth}}
\toprule
\textbf{ID} & \textbf{Questions} & \textbf{Scale} \\
\midrule
Q1 & Which condition was more helpful in cleaning data? & (Condition A, B, C) \\
Q2 & Why did you find it more helpful? & Open-ended \\
Q3 & Why did you find other conditions \textit{less helpful}? & Open-ended \\
Q4 & Any other thoughts, comments, or feedback? & Open-ended \\
\bottomrule
\end{tabular}
\caption{After finishing all the tasks, participants rated which condition was more helpful for data cleaning and provided feedback on why they found certain conditions more or less helpful, using both Likert scale and open-ended responses. (Note: In the survey, UI names were coded to prevent bias)}
\label{tab:poststudy}
\end{table*}



\end{document}